\documentclass[12pt]{article}
\usepackage{amsmath,amssymb}
\usepackage{graphicx}
\newcommand{\eqdef}{\stackrel{\text{def}}{=}}
\newcommand{\n}{\nonumber\\}
\newcommand{\bm}{\boldsymbol}
\newcommand{\ignore}[1]{}
\renewcommand{\theequation}{\arabic{section}.\arabic{equation}}
\newcommand{\Romannumeral}[1]{\uppercase\expandafter{\romannumeral#1}}

\allowdisplaybreaks[4]

\begin{document}

\baselineskip=20pt

\newfont{\elevenmib}{cmmib10 scaled\magstep1}
\newcommand{\preprint}{
    \begin{flushleft}
     \elevenmib Yukawa\, Institute\, Kyoto\\
   \end{flushleft}\vspace{-1.3cm}
   \begin{flushright}\normalsize \sf
     DPSU-11-1\\
     YITP-11-7\\
   \end{flushright}}
\newcommand{\Title}[1]{{\baselineskip=26pt
   \begin{center} \Large \bf #1 \\ \ \\ \end{center}}}
\newcommand{\Author}{\begin{center}
   \large \bf Satoru Odake${}^a$ and Ryu Sasaki${}^b$ \end{center}}
\newcommand{\Address}{\begin{center}
     $^a$ Department of Physics, Shinshu University,\\
     Matsumoto 390-8621, Japan\\
     ${}^b$ Yukawa Institute for Theoretical Physics,\\
     Kyoto University, Kyoto 606-8502, Japan
   \end{center}}
\newcommand{\Accepted}[1]{\begin{center}
   {\large \sf #1}\\ \vspace{1mm}{\small \sf Accepted for Publication}
   \end{center}}

\preprint
\thispagestyle{empty}

\Title{Dual Christoffel transformations}

\Author

\Address
\vspace{1cm}

\begin{abstract}
Crum's theorem and its modification \`a la Krein-Adler are formulated for
the discrete quantum mechanics with real shifts, whose eigenfunctions
consist of orthogonal polynomials of a discrete variable.
The modification produces the associated polynomials with a finite
number of {\em degrees} deleted. This in turn provides the well known
Christoffel transformation for the dual orthogonal polynomials with the
corresponding {\em positions} deleted.
\end{abstract}

\section{Introduction}
\label{intro}

In a previous paper \cite{os12}  we have developed a new paradigm for the
orthogonal polynomials of a discrete variable \cite{nikiforov,askey,ismail,
koeswart}, which is called {\em the discrete quantum mechanics with real
shifts\/}. Various well-known orthogonal polynomials, for example, the
($q$)-Racah, ($q$)-(dual)-Hahn, ($q$)-Meixner, etc.\ \cite{koeswart} are
derived in \cite{os12} as the main parts of the eigenvectors of specific
{\em tri-diagonal\/} real symmetric (Jacobi) matrices of finite or
infinite dimensions. The orthogonality of the eigenpolynomials is evident
by construction and the orthogonality weight function is obtained as the
square of the groundstate eigenvector.
The dual pair of orthogonal polynomials ({\em the Leonard pair\/})
\cite{leonard,terw} is naturally defined and the {\em three term recurrence
relation\/} for the dual polynomials is built in automatically.
It should be emphasised that the present paradigm offers a unified
understanding \cite{os13,os14} of the ($q$)-Askey scheme of hypergeometric
orthogonal polynomials \cite{askey,koeswart}. Various properties of the
discrete quantum mechanical systems with real shifts, for example,
the duality, the orthogonality, the exact solvability in the Schr\"odinger
as well as the Heisenberg picture \cite{os7}, the dynamical symmetry
algebras (the Askey-Wilson algebra \cite{zhedanov} and its degenerations
\cite{os11}),
the quasi-exact solvability, etc. can be understood in a unified fashion,
not one by one specifically.

In the present paper we start with the formulation of the discrete
quantum mechanics version of the well-known theorem by Crum \cite{crum}
for one-dimensional quantum mechanics or Sturm-Liouville theory.
Crum's seminal paper asserts, in the language of quantum mechanics,
the existence of an infinite family of associated Hamiltonian systems
({\it i.e.} the Hamiltonian together with its eigenvalues and
eigenfunctions) which are essentially {\em iso-spectral\/} with each other.
We will show that the situation is the same in the discrete quantum
mechanics with real shifts. The associated Hamiltonians are tri-diagonal
real symmetric (Jacobi) matrices, too.
See Fig.\,1 for the general structure of the associated Hamiltonian systems. 
Crum's theorem for {\em the discrete quantum mechanics with pure imaginary
shifts\/} \cite{os4,os6,os13} has been formulated in \cite{os15}.
This covers the orthogonal polynomials with absolutely continuous weight
functions, for example, the Wilson and the Askey-Wilson polynomials and
their various degenerate forms. See also \cite{matveev} in this context.

Next we will present the discrete quantum mechanics (with real shifts)
version of Krein-Adler \cite{adler} modification of Crum's theorem \cite{gos}.
The Krein-Adler modification goes as follows: starting from a given
Hamiltonian with eigenvalues $\{\mathcal{E}(n)\}$ and the corresponding
eigenfunctions $\{\phi_n\}$, $n=0,1,\ldots$, one can again construct an
associated Hamiltonian system by {\em deleting\/} a finite number of
energy levels $\{\mathcal{E}(d_j)\}$, $d_j\in\mathbb{Z}_{\ge0}$,
$j=1,\ldots,\ell$, which satisfy certain conditions \eqref{dellcond}.
We will demonstrate that the situation is the same in the discrete quantum
mechanics. The associated Hamiltonian is again a tri-diagonal real
symmetric (Jacobi) matrix and the eigenfunctions form a complete set of
orthogonal functions.
See Fig.\,2a for its schematic structure to be contrasted with the original
Crum's case in Fig.\,2b, which corresponds to the very special choice of
deletion $\{d_1,d_2,\ldots,d_{\ell}\}=\{0,1,2,\ldots,\ell-1\}$.
If the starting system is exactly solvable, as those examples given in
\cite{os12}, the modified system is also exactly solvable.

Crum's theorem and its modification will be presented generically, at
the beginning. That is, no extra condition is imposed on the functions
$B(x)$ and $D(x)$ in the Hamiltonian, other than the positivity, the
boundary (asymptotic) conditions \eqref{BDcondition}.
Many formulas will be drastically simplified when the eigenfunctions consist
of polynomials and/or the system is shape invariant \cite{genden}, which
is the case for most practical applications. We will present these
simplified formulas, too.
Due to the lack of the generic `oscillation theorem' in discrete quantum
mechanics, we do not have categorical proof of the hermiticity of the
Hamiltonian $\bar{\mathcal H}$ \eqref{dbfQMHb1bsdef} obtained by the
application of the modified version of Crum's theorem.
For each particular case of specific polynomials the hermiticity is verified.

It is important to stress that the modified Crum's theorem provides a
unified theory of {\em dual Christoffel transformations\/} \cite{chris}
for orthogonal polynomials of a discrete variable.
The dual Christoffel transformation has the merits that the formulas
are universal, concise and algorithmic compared with the original Christoffel
transformation, which should be performed specifically for each case,
the polynomials and the set of deletion $\mathcal{D}$.
This situation is explained in some detail in section \ref{sec:dualChris}.

As an illustration of the modified Crum's theorem, we will derive in Appendix
unified expressions of the eigenpolynomials for the special case of
deletion $\{d_1,d_2,\ldots,d_{\ell}\}=\{1,2,\ldots,\ell\}$.
The same formulation and examples for the discrete quantum mechanics
with pure imaginary shifts are reported in \cite{gos}.
As is well known these orthogonal polynomials of a discrete variable
have many important applications in many arenas of physics and mathematics
\cite{nikiforov,askey,ismail}.
To name a few recent applications, the linear quantum registers \cite{albanese},
quantum communications \cite{chakr} and the birth and death processes.
As shown in \cite{bdproc}, the explicit examples of 18 orthogonal
polynomials in \cite{os12}, the ($q$)-Racah, ($q$)-(dual)-Hahn etc,
provide {\em exactly solvable birth and death processes\/} \cite{ismail,bdp}.
That is, for the given birth and death rates $\{B(x),D(x)\}$ which define
the Hamiltonian \eqref{genham}, the corresponding transition probabilities
are given explicitly, not in a general spectral representation form of
Karlin-McGregor \cite{karlin}.
By applying the present modification of Crum's theorem to these polynomials,
one can generate an (in)finite variety of exactly solvable birth and death
processes.

This article is organised as follows.
In section \ref{sec:dQM} we recapitulate the essence of the discrete
quantum mechanics with real shifts in order to introduce necessary
notions and notation. In section \ref{sec:Crum}, the discrete quantum
mechanics version of Crum's theorem is formulated in its full generality.
The modification of Crum's theorem \`a la Krein-Adler is developed in section
\ref{sec:Adler}. Section \ref{sec:simplifications} provides various
simplifications of the formulas in the presence of shape invariance and/or
polynomial eigenfunctions.
In section \ref{sec:dualChris}, the Christoffel transformations for orthogonal
polynomials of a discrete variable in discrete quantum mechanics
are shown to be {\em dual} to
the modification of Crum's theorem developed in the preceding sections.
The final section is for a summary and comments.
Appendix gives the simplest examples of the modified Hamiltonian systems
obtained by deleting the lowest lying $\ell$ excited states for various
exactly solvable Hamiltonians discussed in \cite{os12}.

\section{Discrete Quantum Mechanics with Real Shifts}
\label{sec:dQM}
\setcounter{equation}{0}

Let us review the discrete quantum mechanics with real shifts formulated in
\cite{os12}.
The Hamiltonian $\mathcal{H}=(\mathcal{H}_{x,y})$ is a very special type of
{\em tri-diagonal\/} real symmetric (Jacobi) matrices and its rows and
columns are indexed by non-negative integers $x$ and $y$
($x,y=0,1,\ldots,x_{\text{max}}$), in which $x_{\text{max}}$ is either finite
($x_{\text{max}}=N$) or infinite ($x_{\text{max}}=\infty$).
The Hamiltonian $\mathcal{H}$ has a form
\begin{align}
  &\mathcal{H}\eqdef
  -\sqrt{B(x)}\,e^{\partial}\sqrt{D(x)}-\sqrt{D(x)}\,e^{-\partial}\sqrt{B(x)}
  +B(x)+D(x),
  \label{genham}\\
  &\mathcal{H}_{x,y}=
  -\sqrt{B(x)D(x+1)}\,\delta_{x+1,y}-\sqrt{B(x-1)D(x)}\,\delta_{x-1,y}
  +\bigl(B(x)+D(x)\bigr)\delta_{x,y},
\end{align}
in which $\partial=\frac{d}{dx}$ is the differentiation operator
$(\partial f)(x)=\frac{df(x)}{dx}$, $(e^{\pm\partial}f)(x)=f(x\pm1)$.
The two functions $B(x)$ and $D(x)$ are real and {\em positive}
but vanish at the boundary:
\begin{align}
  B(x)>0,\quad D(x)>0,\quad D(0)=0\ ;\quad
  B(x_{\text{max}})=0\ \ \text{for the finite case}.
  \label{BDcondition}
\end{align}
The problem at hand is to find the {\em complete set of eigenvalues\/}
and the {\em corresponding eigenvectors\/} of the hermitian matrix
$\mathcal{H}$ ($n_{\text{max}}=N$ or $\infty$):
\begin{equation}
  \mathcal{H}\phi_n(x)=\mathcal{E}(n)\phi_n(x)\quad
  (n=0,1,\ldots,n_{\text{max}}),
\end{equation}
which is the Schr\"{o}dinger equation of the discrete quantum mechanics
with real shifts. It has non-degenerate spectrum thanks to the generic
property of Jacobi matrices,
\begin{equation}
  \mathcal{E}(0)<\mathcal{E}(1)<\mathcal{E}(2)<\cdots.
  \label{Hspec}
\end{equation}
The Hamiltonian \eqref{genham} can be expressed as a product of a
bi-diagonal {\em lower triangular\/} matrix $\mathcal{A}^{\dagger}$
and a bi-diagonal {\em upper triangular\/} matrix $\mathcal{A}$:
\begin{align}
  &\mathcal{H}=\mathcal{A}^{\dagger}\mathcal{A},\qquad
  \mathcal{A}=(\mathcal{A}_{x,y}),\quad
  \mathcal{A}^{\dagger}=((\mathcal{A}^{\dagger})_{x,y})
  =(\mathcal{A}_{y,x})\quad(x,y=0,1,\ldots,x_{\text{max}}),
  \label{factor}\\
  &\mathcal{A}\eqdef\sqrt{B(x)}-e^{\partial}\sqrt{D(x)},\quad
  \mathcal{A}^{\dagger}=\sqrt{B(x)}-\sqrt{D(x)}\,e^{-\partial},
  \label{A,Ad}\\
  &\mathcal{A}_{x,y}=
  \sqrt{B(x)}\,\delta_{x,y}-\sqrt{D(x+1)}\,\delta_{x+1,y},\quad
  (\mathcal{A}^{\dagger})_{x,y}=
  \sqrt{B(x)}\,\delta_{x,y}-\sqrt{D(x)}\,\delta_{x-1,y}.
\end{align}
The zero mode $\mathcal{A}\phi_0(x)=0$  is easily obtained:
\begin{equation}
  \phi_0(x)=\sqrt{\prod_{y=0}^{x-1}\frac{B(y)}{D(y+1)}},
  \label{phi0=prodB/D}
\end{equation}
with the normalisation $\phi_0(0)=1$ (convention: $\prod_{k=n}^{n-1}*=1$).
Note that the groundstate wavefunction $\phi_0(x)$ of \eqref{genham} is
positive ($\phi_0(x)>0$) throughout the range of $x$
($x=0,1,\ldots,x_{\text{max}}$).
 Then the Hamiltonian is {\em positive semi-definite}:
\begin{equation}
  \mathcal{E}(0)=0.
  \label{ezero}
\end{equation}
The finite $\ell^2$-norm condition of $\phi_0(x)$
\begin{equation}
  \sum_{x=0}^{x_{\text{max}}}\phi_0(x)^2<\infty,
  \label{l2norm}
\end{equation}
imposes conditions on the asymptotic forms of $B(x)$ and $D(x)$ for
the infinite case.

The orthogonality relation is
\begin{equation}
  (\phi_n,\phi_m)\eqdef
  \sum_{x=0}^{x_{\text{max}}}\phi_n(x)\phi_m(x)
  =\frac{1}{d_n^2}\,\delta_{nm}\quad
  (n,m=0,1,\ldots,n_{\text{max}}).
  \label{ortho}
\end{equation}
Here $1/d_n^2$ is the normalisation constant.
For the infinite case ($x_{\text{max}}=\infty$), the eigenfunctions should
satisfy the asymptotic condition $\phi_n(x_{\text{max}})=0$ and the finite
$\ell^2$-norm condition.

If the eigenfunction has the factorised form,
\begin{equation}
  \phi_n(x)=\phi_0(x)P_n(\eta(x)),
  \label{phin=phi0Pn}
\end{equation}
where $P_n(\eta)$ is a polynomial of degree $n$ in the sinusoidal coordinate
$\eta=\eta(x)$ with the normalisation,
\begin{equation}
  P_n(0)=1\quad(n=0,1,\ldots,n_{\text{max}}),\quad
  P_{-1}(\eta)\eqdef 0,\quad\eta(0)=0.
  \label{Pnorm}
\end{equation}
The sinusoidal coordinate $\eta(x)$ is the central notion of exactly
solvable quantum mechanics \cite{os12,os13,os7}. It undergoes sinusoidal
motion through the time evolution governed by the Hamiltonian $\mathcal{H}$,
see (2.28) of \cite{os7}. It is also established that the Heisenberg
operator equation for the sinusoidal coordinate $\eta(x)$ is exactly
solvable \cite{os12,os13,os7}.
In the discrete quantum mechanics with real shifts, there are five
sinusoidal coordinates, see (4.72)--(4.76) of \cite{os12}.
Then the above orthogonality relation becomes that of the orthogonal
polynomials
\begin{equation}
  \sum_{x=0}^{x_{\text{max}}}\phi_0(x)^2P_n(\eta(x))P_m(\eta(x))
  =\frac{1}{d_n^2}\,\delta_{nm}\quad
  (n,m=0,1,\ldots,n_{\text{max}}).
  \label{orthP}
\end{equation}
Here the discrete weightfunction $\phi_0(x)^2$ is given explicitly
\eqref{phi0=prodB/D}.
The orthogonal polynomials $P_n(\eta(x))$ are the `eigenfunctions' of the
similarity transformed Hamiltonian
$\widetilde{\mathcal{H}}=(\widetilde{\mathcal{H}}_{x,y})$
($x,y=0,1,\ldots,x_{\text{max}}$)
\begin{align}
  &\widetilde{\mathcal{H}}\eqdef
  \phi_0(x)^{-1}\circ\mathcal{H}\circ\phi_0(x)
  =B(x)(1-e^{\partial})+D(x)(1-e^{-\partial}),\\
  &\widetilde{\mathcal{H}}_{x,y}=
  B(x)(\delta_{x,y}-\delta_{x+1,y})+D(x)(\delta_{x,y}-\delta_{x-1,y}),\\
  &\widetilde{\mathcal{H}}P_n(\eta(x))=\mathcal{E}(n)P_n(\eta(x))
  \quad(n=0,1,\ldots,n_{\text{max}})
  \label{difeqP}\\
  &=B(x)\bigl(P_n(\eta(x))-P_n(\eta(x+1))\bigr)
  +D(x)\bigl(P_n(\eta(x))-P_n(\eta(x-1))\bigr).
  \nonumber
\end{align}

The {\em dual polynomial\/} $Q_x(\mathcal{E})$, which is a degree $x$
polynomial in $\mathcal{E}$ with the normalisation condition
\begin{equation}
  Q_x(0)=1\quad(x=0,1,\ldots,x_{\text{max}}),\quad
  Q_{-1}(\mathcal{E})\eqdef 0,\quad
  \mathcal{E}(0)=0,
\end{equation}
is defined by the {\em three term recurrence relation}
\begin{equation}
  B(x)\bigl(Q_x(\mathcal{E})-Q_{x+1}(\mathcal{E})\bigr)
  +D(x)\bigl(Q_x(\mathcal{E})-Q_{x-1}(\mathcal{E})\bigr)
  =\mathcal{E}Q_x(\mathcal{E})\quad(x=0,1,\ldots,x_{\text{max}}).
\end{equation}
For $\mathcal{E}=\mathcal{E}(n)$, the above difference equation for
$P_n(\eta(x))$ \eqref{difeqP} is identical with the three term recurrence
relation under the identification
\begin{equation}
  P_n(\eta(x))=Q_x(\mathcal{E}(n))\quad(x=0,1,\ldots,x_{\text{max}}),
  \quad(n=0,1,\ldots,n_{\text{max}}).
  \label{duality}
\end{equation}
This establishes the {\em duality} \cite{leonard,terw}.
The orthogonality relation for the dual orthogonal polynomials
$Q_x(\mathcal{E}(n))$ takes a dual form to \eqref{orthP}
\begin{equation}
  \sum_{n=0}^{n_{\text{max}}}d_n^2Q_x(\mathcal{E}(n))Q_y(\mathcal{E}(n))
  =\frac{1}{\phi_0(x)^2}\,\delta_{xy}\quad
  (x,y=0,1,\ldots,x_{\text{max}}).
  \label{orthoQ}
\end{equation}
For more details of the duality in the context of discrete quantum mechanics
with real shifts, see \S\,3 of \cite{os12}.

For all the examples presented in \cite{os12}, the eigenfunctions have
the form \eqref{phin=phi0Pn} and the square of the groundstate wavefunction
$\phi_0(x)^2$ can be analytically continued to the whole complex $x$ plane
and it has (at least) a simple zero at integral points outside
$[0,x_{\text{max}}]$,
\begin{equation}
  \phi_0(x)^2=0 \quad
  \text{for $x\in\mathbb{Z}\backslash\{0,1,\ldots,x_{\text{max}}\}$},
\end{equation}
due to the factors of $\phi_0(x)^2$ such as
$(q\,;q)_x^{-1}=(q^{1+x};q)_{\infty}(q\,;q)_{\infty}^{-1}$, $(q^{-N};q)_x$,
etc. Thanks to this situation, various expressions in the subsequent
sections whose arguments appear to go beyond the defined range of
$[0,x_{\text{max}}]$ cause no harm.

\section{Crum's Theorem}
\label{sec:Crum}
\setcounter{equation}{0}

Crum's theorem \cite{crum}  describes the relationship between the
original and the associated Hamiltonian systems, which are iso-spectral
except for the lowest energy state. The relationship among various
associated Hamiltonian systems is quite general, as depicted in Fig.\,1,
and it is shared by the most general ordinary quantum mechanical systems
\cite{crum} as well as discrete quantum mechanical systems with real and
pure imaginary shifts \cite{os15,matveev}.
As shown below, the factorised form of the Hamiltonian \eqref{factor}
is essential.

\subsection{Deletion of the ground state}
\label{sec:Crum:1st}

For later convenience, let us attach the superscript ${}^{[0]}$ to
all the quantities of the original Hamiltonian system,
$\mathcal{H}^{[0]}\eqdef\mathcal{H}$, $\phi^{[0]}_n(x)\eqdef\phi_n(x)$,
$\mathcal{A}^{[0]}\eqdef\mathcal{A}$,
$\mathcal{A}^{[0]\dagger}\eqdef\mathcal{A}^{\dagger}$,
$B^{[0]}(x)\eqdef B(x)$, $D^{[0]}(x)\eqdef D(x)$.
Let us define an associated Hamiltonian $\mathcal{H}^{[1]}$ by
simply changing the order of $\mathcal{A}$ and $\mathcal{A}^{\dagger}$:
\begin{equation}
  \mathcal{H}^{[1]}\eqdef\mathcal{A}^{[0]}\mathcal{A}^{[0]\dagger}.
  \label{H1def}
\end{equation}
By construction $\mathcal{A}$ and $\mathcal{A}^\dagger$ intertwine
$\mathcal{H}^{[0]}$ and $\mathcal{H}^{[1]}$:
\begin{equation}
  \mathcal{A}^{[0]}\mathcal{H}^{[0]}=\mathcal{H}^{[1]}\mathcal{A}^{[0]},
  \qquad
  \mathcal{A}^{[0]\dagger}\mathcal{H}^{[1]}
  =\mathcal{H}^{[0]}\mathcal{A}^{[0]\dagger}.
  \label{intertwine}
\end{equation}
The matrix elements of the associated Hamiltonian $\mathcal{H}^{[1]}$ are
\begin{equation}
  \mathcal{H}^{[1]}_{x,y}=
  -\sqrt{B(x+1)D(x+1)}\,\delta_{x+1,y}-\sqrt{B(x)D(x)}\,\delta_{x-1,y}
  +\bigl(B(x)+D(x+1)(1-\delta_{x,x_{\text{max}}})\bigr)\delta_{x,y}.
\end{equation}
For the finite case ($x_{\text{max}}^{[0]}=N$, $B(N)=0$), the matrix
elements in the $(N+1)$-st entry vanish, namely $\mathcal{H}^{[1]}$ has
the following form:
\begin{equation}
  \mathcal{H}^{[1]}=
  \begin{pmatrix}\ \mathcal{H}^{[1]\prime}&\bm{0}\\
  {}^t\bm{0}&0
  \end{pmatrix},
\end{equation}
where $\mathcal{H}^{[1]\prime}$ is an $N\times N$ matrix and $\bm{0}$
is an $N$-dimensional zero column vector.
Therefore the eigenvalue problem of $\mathcal{H}^{[1]}$ reduces to that
of $\mathcal{H}^{[1]\prime}$ and the trivial one dimensional part with zero
eigenvalue corresponding to the deleted groundstate of the original
Hamiltonian. To get rid of  this trivial part we define
$x_{\text{max}}^{[1]}$ as $x_{\text{max}}^{[1]}=N-1$.
For the infinite case ($x_{\text{max}}=\infty$), $x_{\text{max}}^{[1]}$ is
defined as $x_{\text{max}}^{[1]}=\infty$.
To treat the finite and the infinite cases in parallel and to avoid
notational cumbersomeness, we write $\mathcal{H}^{[1]\prime}$ as
$\mathcal{H}^{[1]}$ by adopting the following convention.
Namely $\mathcal{H}^{[1]}$ represents an
$(x_{\text{max}}^{[1]}+1)\times(x_{\text{max}}^{[1]}+1)$ matrix
$\mathcal{H}^{[1]}=(\mathcal{H}^{[1]}_{x,y})$
$(x,y=0,1,\ldots,x_{\text{max}}^{[1]}$)
with $x_{\text{max}}^{[1]}=N-1$ or $\infty$.

We will show that the associated Hamiltonian system $\mathcal{H}^{[1]}$
is {\em iso-spectral\/} to the original Hamiltonian system $\mathcal{H}^{[0]}$
and the eigenfunctions are in one to one correspondence, {\em except for\/}
the groundstate.
Thanks to the first of the above relation \eqref{intertwine}, it is trivial
to verify that the eigenfunctions of the associated Hamiltonian system
$\mathcal{H}^{[1]}$ are generated algebraically by multiplying
$\mathcal{A}^{[0]}$ to the eigenfunction of the original system:
\begin{alignat}{2}
  &\phi^{[1]}_n(x)\eqdef\mathcal{A}^{[0]}\phi^{[0]}_n(x)\quad
  &(n=1,2,\ldots,n_{\text{max}}),\\
  &\mathcal{H}^{[1]}\phi^{[1]}_n(x)=\mathcal{E}(n)\phi^{[1]}_n(x)\quad
  &(n=1,2,\ldots,n_{\text{max}}).
  \label{H1spec}
\end{alignat}
The orthogonality relation is
\begin{align}
  (\phi_n^{[1]},\phi_m^{[1]})&\eqdef\sum_{x=0}^{x_{\text{max}}^{[1]}}
  \phi_n^{[1]}(x)\phi_m^{[1]}(x)
  \qquad (n,m=1,2,\ldots,n_{\text{max}})\n
  &=(\mathcal{A}^{[0]}\phi_n^{[0]},\mathcal{A}^{[0]}\phi_m^{[0]})
  =(\phi_n^{[0]},\mathcal{A}^{[0]\dagger}\mathcal{A}^{[0]}\phi_m^{[0]})
  =(\phi_n^{[0]},\mathcal{H}^{[0]}\phi_m^{[0]})
  =\mathcal{E}(n)\frac{1}{d_n^2}\,\delta_{nm}.
\end{align}
For the finite case this gives all the $N$ eigenvectors of $\mathcal{H}^{[1]}$.
For the infinite case, suppose the associated Hamiltonian $\mathcal{H}^{[1]}$
has an eigenfunction $\phi'(x)$ with an eigenvalue $\mathcal{E}'$
other than those listed above:
\begin{equation}
  \mathcal{H}^{[1]}\phi'(x)=\mathcal{E}'\phi'(x).
\end{equation}
Again, thanks to the second of the relation \eqref{intertwine},
it is trivial to verify
\begin{equation}
  \mathcal{H}^{[0]}\mathcal{A}^{[0]\dagger}\phi'(x)
  =\mathcal{E}'\mathcal{A}^{[0]\dagger}\phi'(x).
\end{equation}
Due to the {\em completeness\/} of the spectrum of the original Hamiltonian
$\mathcal{H}^{[0]}$, the provisional eigenvalue $\mathcal{E}'$ must
belong to the spectrum \eqref{Hspec} $\mathcal{E}(n)$ for
$n=1,2.\ldots,n_{\text{max}}$.
In other words, $\mathcal{E}'$ cannot be vanishing, $\mathcal{E}'\neq0$.
Suppose that is the case ($\mathcal{E}'=0$), then $\phi'$ is annihilated by
$\mathcal{A}^{[0]\dagger}$.
But it is easy to see that there exists no non-zero solution (the finite
case) or no finite $\ell^2$-norm solution (the infinite case) of the equation
$\mathcal{A}^{[0]\dagger}\phi'(x)=0$.
Thus we have established that the associated Hamiltonian system
$\mathcal{H}^{[1]}$ is essentially {\em iso-spectral\/} to the original
Hamiltonian system $\mathcal{H}^{[0]}$ and the eigenfunctions are in one
to one correspondence, {\em except for\/} the groundstate with the
wavefunction $\phi_0^{[0]}(x)$.

If the groundstate energy $\mathcal{E}(1)$ is subtracted from the
associated Hamiltonian $\mathcal{H}^{[1]}$, it is again positive
semi-definite and can be factorised as above:
\begin{align}
  &\mathcal{H}^{[1]}=\mathcal{A}^{[1]\dagger}\mathcal{A}^{[1]}+\mathcal{E}(1),
  \label{H1rewrite}\\
  &\mathcal{A}^{[1]}\eqdef\sqrt{B^{[1]}(x)}-e^{\partial}\sqrt{D^{[1]}(x)},
  \quad
  \mathcal{A}^{[1]\dagger}=
  \sqrt{B^{[1]}(x)}-\sqrt{D^{[1]}(x)}\,e^{-\partial},
  \label{A1}\\
  &B^{[1]}(x)\eqdef\sqrt{B^{[0]}(x+1)D^{[0]}(x+1)}\,
  \frac{\phi_1^{[1]}(x+1)}{\phi_1^{[1]}(x)},
  \label{B1}\\
  &D^{[1]}(x)\eqdef\sqrt{B^{[0]}(x)D^{[0]}(x)}\,
  \frac{\phi_1^{[1]}(x-1)}{\phi_1^{[1]}(x)}.
  \label{D1}
\end{align}
Since $\phi_1^{[1]}$ is the groundstate of $\mathcal{H}^{[1]}$, it does
not vanish inside the interval $[0,x_{\text{max}}^{[1]}]$.
Thus $B^{[1]}(x)$ and $D^{[1]}(x)$ are non-singular and positive.
Note that $D^{[1]}(0)$ is not yet defined because $\phi_1^{[1]}(-1)$
is not defined. Due to the factor $D^{[0]}(x)$ in \eqref{D1} and property
$D^{[0]}(0)=0$, we define $D^{[1]}(0)\eqdef 0$.
In concrete examples the expression of $\phi_1^{[1]}(x)^2$ can be analytically
continued and it vanishes at $x=-1$.
This fact also supports the definition $D^{[1]}(0)=0$.
The groundstate wavefunction $\phi_1^{[1]}(x)$ has the following form
\begin{equation}
  \phi_1^{[1]}(x)=\mathcal{A}^{[0]}\phi_1^{[0]}(x)
  =\sqrt{B^{[0]}(x)}\,\phi_1^{[0]}(x)
  -\sqrt{D^{[0]}(x+1)}\,\phi_1^{[0]}(x+1)(1-\delta_{x,x_{\text{max}}^{[0]}}).
\end{equation}
Due to the vanishing factor $B^{[0]}(x_{\text{max}}^{[0]})=0$ for the
finite case and $\phi^{[0]}_n(x_{\text{max}}^{[0]})=0$ for the infinite
case, this expression of the groundstate wavefunction $\phi_1^{[1]}(x)$
vanishes at $x=x_{\text{max}}^{[0]}$,
\begin{equation}
  \phi_1^{[1]}(x_{\text{max}}^{[0]})=0.
\end{equation}
Thus we have $B^{[1]}(x_{\text{max}}^{[1]})=0$ for the finite case.

It is easy to verify that $\phi_1^{[1]}(x)$ is annihilated by
$\mathcal{A}^{[1]}$,
\begin{equation}
  \mathcal{A}^{[1]}\phi_1^{[1]}(x)=0.
\end{equation}
The equation \eqref{H1rewrite} is shown by elementary calculation.

\subsection{Repetition}
\label{sec:Crum:repeat}

Starting from \eqref{H1rewrite}, the second associated Hamiltonian system
$\mathcal{H}^{[2]}$ can be defined by reversing the order of
$\mathcal{A}^{[1]\dagger}$ and $\mathcal{A}^{[1]}$.
This process can go on repeatedly.

Here we list the definition of the $s$-th quantities step by step
for $s\geq 1$,
\begin{align}
  &\mathcal{H}^{[s]}\eqdef\mathcal{A}^{[s-1]}\mathcal{A}^{[s-1]\,\dagger}
  +\mathcal{E}(s-1),\quad
  x_{\text{max}}^{[s]}\eqdef N-s \text{ or } \infty,\\
  &\phi^{[s]}_n(x)\eqdef\mathcal{A}^{[s-1]}\phi^{[s-1]}_n(x)\quad
  (n=s,s+1,\ldots,n_{\text{max}}),\\
  &\mathcal{A}^{[s]}\eqdef\sqrt{B^{[s]}(x)}-e^{\partial}\sqrt{D^{[s]}(x)},\quad
  \mathcal{A}^{[s]\dagger}=
  \sqrt{B^{[s]}(x)}-\sqrt{D^{[s]}(x)}\,e^{-\partial},\\
  &B^{[s]}(x)\eqdef\sqrt{B^{[s-1]}(x+1)D^{[s-1]}(x+1)}\,
  \frac{\phi_s^{[s]}(x+1)}{\phi_s^{[s]}(x)},\\
  &D^{[s]}(x)\eqdef\sqrt{B^{[s-1]}(x)D^{[s-1]}(x)}\,
  \frac{\phi_s^{[s]}(x-1)}{\phi_s^{[s]}(x)}.
\end{align}
Like in the $s=1$ case we set $D^{[s]}(0)=0$.
Recall that $\mathcal{H}^{[s]}$ is an
$(x_{\text{max}}^{[s]}+1)\times(x_{\text{max}}^{[s]}+1)$ matrix
$\mathcal{H}^{[s]}=(\mathcal{H}^{[s]}_{x,y})$
$(x,y=0,1,\ldots,x_{\text{max}}^{[s]}$).
Then we can show the following for $s\geq 0$,
\begin{align}
  &\mathcal{H}^{[s]}\phi^{[s]}_n(x)=\mathcal{E}(n)\phi^{[s]}_n(x)\quad
  (n=s,s+1,\ldots,n_{\text{max}}),\\
  &\mathcal{A}^{[s]}\phi^{[s]}_s(x)=0,\\
  &\mathcal{H}^{[s]}=\mathcal{A}^{[s]\,\dagger}\mathcal{A}^{[s]}
  +\mathcal{E}({s}),\\
  &(\phi_n^{[s]},\phi_m^{[s]})\eqdef\sum_{x=0}^{x_{\text{max}}^{[s]}}
  \phi_n^{[s]}(x)\phi_m^{[s]}(x)
  =\prod_{j=0}^{s-1}(\mathcal{E}(n)-\mathcal{E}(j))\cdot\frac{1}{d_n^2}\,
  \delta_{nm}
  \quad (n,m=s,s+1,\ldots,n_{\text{max}}),
\end{align}
and $B^{[s]}(x_{\text{max}}^{[s]})=0$ for the finite case.
We have also for $s\geq 1$
\begin{equation}
  \phi^{[s-1]}_n(x)=\frac{\mathcal{A}^{[s-1]\,\dagger}}
  {\mathcal{E}(n)-\mathcal{E}({s-1})}\,\phi^{[s]}_n(x)\quad
  (n=s,s+1,\ldots,n_{\text{max}}).
\end{equation}
The situation of Crum's theorem is illustrated in Fig.\,1.

As in the original Crum's case \cite{crum}, the eigenfunction
$\phi_n^{[s]}(x)$ can be expressed in terms of determinants.
Let us define the Casorati determinant for $n$ functions $f_j(x)$ as
\begin{equation}
  \text{W}[f_1,\ldots,f_n](x)
  \eqdef\det\Bigl(f_k(x+j-1)\Bigr)_{1\leq j,k\leq n},
\end{equation}
(for $n=0$, we set $\text{W}[\cdot](x)=1$), which satisfies
\begin{align}
  &\text{W}[gf_1,gf_2,\ldots,gf_n](x)
  =\prod_{k=0}^{n-1}g(x+k)\cdot\text{W}[f_1,f_2,\ldots,f_n](x),
  \label{Wformula1}\\
  &\text{W}\bigl[\text{W}[f_1,f_2,\ldots,f_n,g],
  \text{W}[f_1,f_2,\ldots,f_n,h]\,\bigr](x)\n
  &=\text{W}[f_1,f_2,\ldots,f_n](x+1)\,
  \text{W}[f_1,f_2,\ldots,f_n,g,h](x)
  \qquad(n\geq 0).
  \label{Wformula2}
\end{align}
By using the Casorati determinant, we obtain
\begin{align}
  \phi_n^{[s]}(x)&=(-1)^s\prod_{k=0}^{s-1}\sqrt{B^{[k]}(x)}\cdot
  \frac{\text{W}[\phi_0,\phi_1,\ldots,\phi_{s-1},\phi_n](x)}
  {\text{W}[\phi_0,\phi_1,\ldots,\phi_{s-1}](x+1)}\\
  &=(-1)^s\prod_{k=0}^{s-1}\sqrt{D^{[k]}(x+s-k)}\cdot
  \frac{\text{W}[\phi_0,\phi_1,\ldots,\phi_{s-1},\phi_n](x)}
  {\text{W}[\phi_0,\phi_1,\ldots,\phi_{s-1}](x)}.
\end{align}

\begin{center}
  \includegraphics{crumscheme2.epsi}
\end{center}
\begin{center}
  Figure\,1: Schematic picture of Crum's theorem
\end{center}

\section{Adler's Modification of Crum's Theorem}
\label{sec:Adler}
\setcounter{equation}{0}

Crum's theorem describes the construction of an associated Hamiltonian
system which is iso-spectral to the original one with the lowest energy
state deleted.
Adler's modification \cite{adler} of Crum's theorem is the construction
of an associated Hamiltonian system which is iso-spectral to the original
one with finitely many states deleted.

Let us choose a set of $\ell$ distinct non-negative integers\footnote{
Although this notation $d_j$ conflicts with the notation of the normalisation
constant $d_n$ in \eqref{ortho}, we think this does not cause any confusion
because the latter appears as $\frac{1}{d_n^2}\,\delta_{nm}$.
}
$\mathcal{D}\eqdef\{d_1,d_2,\ldots,d_{\ell}\}\subset\mathbb{Z}_{\ge 0}^{\ell}$,
satisfying the condition \cite{adler}
\begin{equation}
  \prod_{j=1}^\ell(m-d_j)\ge0,\quad\forall m\in\mathbb{Z}_{\ge 0}.
  \label{dellcond}
\end{equation}
This condition means that the set $\mathcal{D}$ consists of several
clusters, each containing an {\em even number\/} of {\em contiguous\/}
integers
\begin{equation}
  d_{k_1}, d_{k_1}+1,\cdots,d_{k_2}\ ;\ d_{k_3},
  d_{k_3}+1,\cdots, d_{k_4}\ ;\ d_{k_5},
  d_{k_5}+1,\cdots, d_{k_6}\ ;\ \cdots,
\end{equation}
where $d_{k_2}+1<d_{k_3},\ d_{k_4}+1<d_{k_5},\ \cdots$.
If $d_{k_1}=0$ for the lowest lying cluster, it could contain an even
or odd number of contiguous integers.
The set $\mathcal{D}$ specifies the energy levels to be {\em deleted}.
Deleting an arbitrary number of contiguous energy levels
starting from the groundstate ($\mathcal{D}=\{0,1,2,\ldots,\ell-1\}$) is
achieved by Crum's theorem discussed in \S\,\ref{sec:Crum}.

\bigskip
\begin{figure}[htbp]
\begin{center}
  \includegraphics{adlerschemefig2.epsi}
  \hspace*{20mm}
  \includegraphics{crum2.epsi}
\end{center}
\vspace*{-3mm}
\hspace*{20mm}
Figure\,2a: Generic case
\hspace*{28mm}
Figure\,2b: Crum's case
\end{figure}

We will construct associated Hamiltonian systems corresponding to the
successive deletions $\mathcal{H}_{d_1\ldots}$ (and
$\mathcal{A}_{d_1\ldots}$, $\mathcal{A}_{d_1\ldots}^{\dagger}$, etc.)
step by step, algebraically. It should be noted that some  Hamiltonians
in the intermediate steps could be non-hermitian.

\subsection{First step}
\label{sec:Adler:1st}

For given $d_1$, the original Hamiltonian $\mathcal{H}$ can be expressed
in two different ways:
\begin{align}
  &\mathcal{H}=\mathcal{A}^{\dagger}\mathcal{A}
  =\mathcal{A}^{\dagger}_{d_1}\mathcal{A}_{d_1}
  +\mathcal{E}({d_1}),
  \label{H=Ad1dAd1}\\
  &\mathcal{A}_{d_1}\eqdef\sqrt{B_{d_1}(x)}-e^{\partial}\sqrt{D_{d_1}(x)},
  \quad
  \mathcal{A}_{d_1}^{\dagger}
  \eqdef\sqrt{B_{d_1}(x)}-\sqrt{D_{d_1}(x)}\,e^{-\partial},
  \label{Ad1Ad1d}\\
  &B_{d_1}(x)\eqdef\sqrt{B(x)D(x+1)}\,\frac{\phi_{d_1}(x+1)}{\phi_{d_1}(x)},
  \quad
  D_{d_1}(x)\eqdef\sqrt{B(x-1)D(x)}\,\frac{\phi_{d_1}(x-1)}{\phi_{d_1}(x)},
\end{align}
and we have
\begin{equation}
  \mathcal{A}_{d_1}\phi_{d_1}(x)=0.
\end{equation}
As in \S\,\ref{sec:Crum} we set $D_{d_1}(0)=0$.
Note that $B_{d_1}(x_{\text{max}})=0$ for the finite case.
Unless $d_1=0$, $B_{d_1}(x)$ and $D_{d_1}(x)$ are not always positive
due to the zeros of $\phi_{d_1}(x)$.
It is important to note that $\mathcal{A}_{d_1}^{\dagger}$ in \eqref{Ad1Ad1d}
is a `formal adjoint' of $\mathcal{A}_{d_1}$ due to the above mentioned
sign changes of $B_{d_1}(x)$ and $D_{d_1}(x)$.
We stick to this notation, since the algebraic structure of various
expressions appearing in the deletion processes
are best described by using the `formal adjoint'.
By changing the order of $\mathcal{A}_{d_1}$ and $\mathcal{A}_{d_1}^{\dagger}$,
let us define a new Hamiltonian system
\begin{equation}
  \mathcal{H}_{d_1}\eqdef\mathcal{A}_{d_1}\mathcal{A}_{d_1}^{\dagger}
  +\mathcal{E}({d_1}),\quad
  x_{\text{max}}^{d_1}\eqdef N-1 \text{ or } \infty.
\end{equation}
As in  \S\,\ref{sec:Crum}, the Hamiltonian $\mathcal{H}_{d_1}$ represents
an $(x_{\text{max}}^{d_1}+1)\times(x_{\text{max}}^{d_1}+1)$ matrix
$\mathcal{H}_{d_1}=(\mathcal{H}_{d_1;\,x,y})$
$(x,y=0,1,\ldots,x_{\text{max}}^{d_1}$).
It is easy to show that the `eigenfunctions' of this Hamiltonian are given by
\begin{alignat}{2}
  &\phi_{d_1\, n}(x)\eqdef\mathcal{A}_{d_1}\phi_n(x)\quad
  &(n\in\{0,1,\ldots,n_{\text{max}}\}\backslash\{d_1\}),\\
  &\mathcal{H}_{d_1}\phi_{d_1\,n}(x)=\mathcal{E}(n)\phi_{d_1\,n}(x)\quad
  &(n\in\{0,1,\ldots,n_{\text{max}}\}\backslash\{d_1\}).
\end{alignat}
Thus the energy level $d_1$ is now deleted, $\phi_{d_1\, d_1}(x)\equiv0$,
from the set of `eigenfunctions' $\{\phi_{d_1\, n}(x)\}$ of the new
Hamiltonian $\mathcal{H}_{d_1}$.

\subsection{Repetition}
\label{sec:Adler:repeat}

Suppose we have determined the Hamiltonian $\mathcal{H}_{d_1\,\ldots\,d_s}$
together with the eigenfunctions $\phi_{d_1\,\ldots\,d_s\,n}(x)$ with $s$
deletions. They have the following properties:
\begin{align}
  &\mathcal{H}_{d_1\,\ldots\,d_s}\eqdef\mathcal{A}_{d_1\,\ldots\,d_s}
  \mathcal{A}_{d_1\,\ldots\,d_s}^{\dagger}+\mathcal{E}({d_s}),\quad
  x_{\text{max}}^{d_1\,\ldots\,d_s}\eqdef N-s \text{ or } \infty,
  \label{dQMHb1bsdef}\\
  &\mathcal{A}_{d_1\,\ldots\,d_s}\eqdef
  \sqrt{B_{d_1\,\ldots\,d_s}(x)} -e^{\partial}\sqrt{D_{d_1\,\ldots\,d_s}(x)},\n
  &\mathcal{A}_{d_1\,\ldots\,d_s}^{\dagger}\eqdef
  \sqrt{B_{d_1\,\ldots\,d_s}(x)} -\sqrt{D_{d_1\,\ldots\,d_s}(x)}\,
  e^{-\partial},\\
  &B_{d_1\,\ldots\,d_s}(x)\eqdef
  \left\{\begin{array}{ll}
  {\displaystyle
  \sqrt{B_{d_1\,\ldots\,d_{s-1}}(x+1) D_{d_1\,\ldots\,d_{s-1}}(x+1)}\,
  \frac{\phi_{d_1\,\ldots\,d_s}(x+1)}{\phi_{d_1\,\ldots\,d_s}(x)}}&(s\geq 2)\\
  {\displaystyle
  \sqrt{B(x)D(x+1)}\,\frac{\phi_{d_1}(x+1)}{\phi_{d_1}(x)}}&(s=1)
  \end{array}\right.,\\
  &D_{d_1\,\ldots\,d_s}(x)\eqdef
  \left\{\begin{array}{ll}
  {\displaystyle
  \sqrt{B_{d_1\,\ldots\,d_{s-1}}(x)D_{d_1\,\ldots\,d_{s-1}}(x)}\,
  \frac{\phi_{d_1\,\ldots\,d_s}(x-1)}{\phi_{d_1\,\ldots\,d_s}(x)}}&(s\geq 2)\\
  {\displaystyle
  \sqrt{B(x-1)D(x)}\,\frac{\phi_{d_1}(x-1)}{\phi_{d_1}(x)}}&(s=1)
  \end{array}\right.,\\
  &\phi_{d_1\,\ldots\,d_s\,n}(x)
  \eqdef\mathcal{A}_{d_1\,\ldots\,d_s}\phi_{d_1\,\ldots\,d_{s-1}\,n}(x),
  \label{dQMphib1bsndef}\\
  &\mathcal{H}_{d_1\,\ldots\,d_s}\phi_{d_1\,\ldots\,d_s\,n}(x)
  =\mathcal{E}(n)\phi_{d_1\,\ldots\,d_s\,n}(x),
  \label{dQMHb1bnphi=..}
\end{align}
where $n\in\{0,1,\ldots,n_{\text{max}}\}\backslash\{d_1,\ldots,d_s\}$.
As before we set $D_{d_1\,\ldots\,d_s}(0)=0$.
We also have $B_{d_1\,\ldots\,d_s}(x_{\text{max}}^{[s]})=0$ for the finite case.
We note that the following relations hold:
\begin{align}
  &B_{d_1\,\ldots\,d_s}(x)=D_{d_1\,\ldots\,d_{s}}(x+1)
  \Bigl(\frac{\phi_{d_1\,\ldots\,d_s}(x+1)}{\phi_{d_1\,\ldots\,d_s}(x)}\Bigr)^2
  \quad(s\geq 1),\\
  &B_{d_1\,\ldots\,d_s}(x)D_{d_1\,\ldots\,d_{s}}(x+1)=
  \left\{\begin{array}{ll}
  {\displaystyle
  B_{d_1\,\ldots\,d_{s-1}}(x+1)D_{d_1\,\ldots\,d_{s-1}}(x+1)}&(s\geq 2)\\[2pt]
  {\displaystyle
  B(x)D(x+1)}&(s=1)
  \end{array}\right.,\\
  &B_{d_1\,\ldots\,d_s}(x)+D_{d_1\,\ldots\,d_s}(x)+\mathcal{E}({d_s})=
  \left\{\begin{array}{ll}
  \!\!{\displaystyle
  B_{d_1\,\ldots\,d_{s-1}}(x)+D_{d_1\,\ldots\,d_{s-1}}(x+1)
  +\mathcal{E}({d_{s-1}}})&(s\geq 2)\\[2pt]
  \!\!{\displaystyle
  B(x)+D(x)}&(s=1)
  \end{array}\right.\!\!.
\end{align}
We have also
\begin{equation}
  \phi_{d_1\,\ldots\,d_{s-1}\,n}(x)
  =\frac{\mathcal{A}_{d_1\,\ldots\,d_s}^{\dagger}}
  {\mathcal{E}(n)-\mathcal{E}({d_s})}\phi_{d_1\,\ldots\,d_s\,n}(x)
  \quad(n\in\{0,1,\ldots,n_{\text{max}}\}\backslash\{d_1,\ldots,d_s\}).
\end{equation}

Next we will define a new Hamiltonian system with one more deletion of
the level $d_{s+1}$.
We can show the following:
\begin{align}
  &\mathcal{H}_{d_1\,\ldots\,d_{s}}
  =\mathcal{A}_{d_1\,\ldots\,d_s\,d_{s+1}}^{\dagger}
  \mathcal{A}_{d_1\,\ldots\,d_s\,d_{s+1}}+\mathcal{E}({d_{s+1}}),\quad
  \mathcal{A}_{d_1\,\ldots\,d_s\,d_{s+1}}\phi_{d_1\,\ldots\,d_s\,d_{s+1}}(x)=0,
  \label{Hb1bs}\\
  &\mathcal{A}_{d_1\,\ldots\,d_s\,d_{s+1}}\eqdef
  \sqrt{B_{d_1\,\ldots\,d_s\,d_{s+1}}(x)}
  -e^{\partial}\sqrt{D_{d_1\,\ldots\,d_s\,d_{s+1}}(x)},\n
  &\mathcal{A}_{d_1\,\ldots\,d_s\,d_{s+1}}^{\dagger}\eqdef
  \sqrt{B_{d_1\,\ldots\,d_s\,d_{s+1}}(x)}
  -\sqrt{D_{d_1\,\ldots\,d_s\,d_{s+1}}(x)}\,e^{-\partial},\\
  &B_{d_1\,\ldots\,d_s\,d_{s+1}}(x)\eqdef
  \sqrt{B_{d_1\,\ldots\,d_s}(x+1)D_{d_1\,\ldots\,d_s}(x+1)}\,
  \frac{\phi_{d_1\,\ldots\,d_s\,d_{s+1}}(x+1)}
  {\phi_{d_1\,\ldots\,d_s\,d_{s+1}}(x)},\\
  &D_{d_1\,\ldots\,d_s\,d_{s+1}}(x)\eqdef
  \sqrt{B_{d_1\,\ldots\,d_s}(x)D_{d_1\,\ldots\,d_s}(x)}\,
  \frac{\phi_{d_1\,\ldots\,d_s\,d_{s+1}}(x-1)}
  {\phi_{d_1\,\ldots\,d_s\,d_{s+1}}(x)}.
\end{align}
These determine a new Hamiltonian system with $s+1$ deletions:
\begin{align}
  &\mathcal{H}_{d_1\,\ldots\,d_{s+1}}
  \eqdef\mathcal{A}_{d_1\,\ldots\,d_{s+1}}
  \mathcal{A}_{d_1\,\ldots\,d_{s+1}}^{\dagger}+\mathcal{E}({d_{s+1}}),\quad
  x_{\text{max}}^{d_1\,\ldots\,d_{s+1}}\eqdef N-s-1 \text{ or } \infty,
  \label{d2QMHb1bsdef}\\
  &\phi_{d_1\,\ldots\,d_{s+1}\,n}(x)
  \eqdef\mathcal{A}_{d_1\,\ldots\,d_{s+1}}\phi_{d_1\,\ldots\,d_{s}\,n}(x),
  \label{d2QMphib1bsndef}\\
  &\mathcal{H}_{d_1\,\ldots\,d_{s+1}}\phi_{d_1\,\ldots\,d_{s+1}\,n}(x)
  =\mathcal{E}(n)\phi_{d_1\,\ldots\,d_{s+1}\,n}(x),
  \label{d2QMHb1bnphi=..}
\end{align}
where $n\in\{0,1,\ldots,n_{\text{max}}\}\backslash\{d_1,\ldots,d_{s+1}\}$.

By induction we can show that the eigenfunction is expressed in terms
of the Casorati determinant:
\begin{align}
  \phi_{d_1\,\ldots\,d_s\,n}(x)
  &=(-1)^s\sqrt{\prod_{k=1}^s B_{d_1\,\ldots\,d_k}(x)}\cdot
  \frac{\text{W}[\phi_{d_1},\ldots,\phi_{d_s},\phi_n](x)}
  {\text{W}[\phi_{d_1},\ldots,\phi_{d_s}](x+1)}\\
  &=(-1)^s\sqrt{\prod_{k=1}^sD_{d_1\,\ldots\,d_k}(x+s+1-k)}\cdot
  \frac{\text{W}[\phi_{d_1},\ldots,\phi_{d_s},\phi_n](x)}
  {\text{W}[\phi_{d_1},\ldots,\phi_{d_s}](x)}.
\end{align}
We can also show the following:
\begin{align}
  \prod_{k=1}^sB_{d_1\,\ldots\,d_k}(x)&=
  \sqrt{\prod_{k=1}^sB(x+k-1)D(x+k)}\cdot
  \frac{\text{W}[\phi_{d_1},\ldots,\phi_{d_s}](x+1)}
  {\text{W}[\phi_{d_1},\ldots,\phi_{d_s}](x)},
  \label{Bsymp}\\
  \prod_{k=1}^sD_{d_1\,\ldots\,d_k}(x+s+1-k)&=
  \sqrt{\prod_{k=1}^sB(x+k-1)D(x+k)}\cdot
  \frac{\text{W}[\phi_{d_1},\ldots,\phi_{d_s}](x)}
  {\text{W}[\phi_{d_1},\ldots,\phi_{d_s}](x+1)}.
  \label{Dsymp}
\end{align}

\subsection{Last step}
\label{sec:Adler:last}

After deleting all the $\mathcal{D}=\{d_1,\,\cdots,\,d_{\ell}\}$ energy
levels, the resulting Hamiltonian system
$\mathcal{H}_{\mathcal{D}}\equiv\mathcal{H}_{d_1\,\ldots\,d_{\ell}}$,
$\mathcal{A}_{\mathcal{D}}\equiv\mathcal{A}_{d_1\,\ldots\,d_{\ell}}$,
etc has the following form:
\begin{align}
  &\mathcal{H}_{\mathcal{D}}
  \eqdef\mathcal{A}_{\mathcal{D}}\mathcal{A}_{\mathcal{D}}^{\dagger}
  +\mathcal{E}(d_{\ell}),\quad
  x_{\text{max}}^{\mathcal{D}}\eqdef N-\ell \text{ or } \infty,
  \label{2dfQMHb1bsdef}\\
  &\mathcal{A}_{\mathcal{D}}\eqdef
  \sqrt{B_{\mathcal{D}}(x)}-e^{\partial}\sqrt{D_{\mathcal{D}}(x)},\quad
  \mathcal{A}_{\mathcal{D}}^{\dagger}\eqdef
  \sqrt{B_{\mathcal{D}}(x)}-\sqrt{D_{\mathcal{D}}(x)}\,e^{-\partial},\\
  &B_{\mathcal{D}}(x)\eqdef
  \sqrt{B_{d_1\,\ldots\,d_{\ell-1}}(x+1)D_{d_1\,\ldots\,d_{\ell-1}}(x+1)}\,
  \frac{\phi_{\mathcal{D}}(x+1)}{\phi_{\mathcal{D}}(x)},\\
  &D_{\mathcal{D}}(x)\eqdef
  \sqrt{B_{d_1\,\ldots\,d_{\ell-1}}(x)D_{d_1\,\ldots\,d_{\ell-1}}(x)}\,
  \frac{\phi_{\mathcal{D}}(x-1)}{\phi_{\mathcal{D}}(x)},\\
  &\phi_{{\mathcal{D}}\,n}(x)
  \eqdef\mathcal{A}_{\mathcal{D}}\phi_{d_1\,\ldots\,d_{\ell-1}\,n}(x)
  \quad(n\in\{0,1,\ldots,n_{\text{max}}\}\backslash{\mathcal{D}}),
  \label{2dQMphib1bsndef}\\
  &\mathcal{H}_{\mathcal{D}}\phi_{{\mathcal{D}}\,n}(x)
  =\mathcal{E}(n)\phi_{{\mathcal{D}}\,n}(x)
  \quad(n\in\{0,1,\ldots,n_{\text{max}}\}\backslash{\mathcal{D}}).
  \label{2dfQMHb1bnphi=..}
\end{align}
Now $\mathcal{H}_{\mathcal{D}}$ has the lowest energy level $\mu$:
\begin{equation}
  \mu\eqdef \min\{n\,|\,n\in\{0,1,\ldots,n_{\text{max}}\}
  \backslash{\mathcal{D}}\},
  \label{mudef}
\end{equation}
with the groundstate wavefunction $\bar{\phi}_\mu(x)$
\begin{equation}
  \bar{\phi}_\mu(x)\eqdef\phi_{{\mathcal{D}}\,\mu}(x)\equiv
  \phi_{d_1\,\cdots\,d_{\ell}\, \mu}(x).
\end{equation}
Then the Hamiltonian system can be expressed simply in terms of the
groundstate wavefunction $\bar{\phi}_\mu(x)$, which we will denote
by new symbols $\bar{\mathcal{H}}$, $\bar{\mathcal{A}}$, etc:
\begin{align}
  &\bar{\mathcal{H}}\equiv\mathcal{H}_{\mathcal{D}}
  \eqdef\bar{\mathcal{A}}^\dagger\bar{\mathcal{A}}+\mathcal{E}({\mu}),
  \quad
  \bar{x}_{\text{max}}\equiv x_{\text{max}}^{\mathcal{D}}
  =N-\ell \text{ or } \infty,
  \quad
  \bar{\mathcal{A}}\bar{\phi}_{\mu}(x)=0,
  \label{dbfQMHb1bsdef}\\
  &\bar{\mathcal{A}}\equiv \mathcal{A}_{\mathcal{D}\,\mu}\eqdef
  \sqrt{\bar{B}(x)}-e^{\partial}\sqrt{\bar{D}(x)},\quad
  \bar{\mathcal{A}}^{\dagger}\equiv
  \mathcal{A}_{\mathcal{D}\,\mu}^\dagger\eqdef
  \sqrt{\bar{B}(x)}-\sqrt{\bar{D}(x)}\,e^{-\partial},
  \label{dbfQMAAd}\\
  &\bar{B}(x)\equiv B_{\mathcal{D}\,\mu}(x)\eqdef
  \sqrt{B_{\mathcal{D}}(x+1)D_{\mathcal{D}}(x+1)}\,
  \frac{\bar{\phi}_{\mu}(x+1)}{\bar{\phi}_{\mu}(x)},\\
  &\bar{D}(x)\equiv D_{\mathcal{D}\,\mu}(x)\eqdef
  \sqrt{B_{\mathcal{D}}(x)D_{\mathcal{D}}(x)}\,
  \frac{\bar{\phi}_{\mu}(x-1)}{\bar{\phi}_{\mu}(x)},\\
  &\bar{\mathcal{H}}\bar{\phi}_{n}(x)
  =\mathcal{E}(n)\bar{\phi}_{n}(x),\quad
  \bar{\phi}_{n}(x)\equiv\phi_{{\mathcal{D}}\,n}(x)
  \quad(n\in\{0,1,\ldots,n_{\text{max}}\}\backslash{\mathcal{D}}),
  \label{dbfQMHb1bnphi=..}\\
  &\bar{B}(x)+\bar{D}(x)+\mathcal{E}(\mu)=
  B_{\mathcal{D}}(x)+D_{\mathcal{D}}(x+1)+\mathcal{E}(d_{\ell}).
\end{align}
Again we set $\bar{D}(0)=0$.
We have $\bar{B}(\bar{x}_{\text{max}})=0$ for the finite case.
Note that the final Hamiltonian $\bar{\mathcal{H}}$ is an
$(\bar{x}_{\text{max}}+1)\times(\bar{x}_{\text{max}}+1)$ matrix
$\bar{\mathcal{H}}=(\bar{\mathcal{H}}_{x,y})$
$(x,y=0,1,\ldots,\bar{x}_{\text{max}}$).

In terms of the Casorati determinants we obtain  the expressions of the
final eigenfunction $\bar{\phi}_n(x)$ and the final functions $\bar{B}(x)$
and $\bar{D}(x)$ ($\ell\geq 0$)
\begin{align}
  \bar{\phi}_n(x)&=
  (-1)^{\ell}\sqrt{\prod_{k=1}^{\ell}B_{d_1\,\ldots\,d_k}(x)}\cdot
  \frac{\text{W}[\phi_{d_1},\ldots,\phi_{d_{\ell}},\phi_n](x)}
  {\text{W}[\phi_{d_1},\ldots,\phi_{d_{\ell}}](x+1)}
  \label{phibarnB}\\
  &=(-1)^{\ell}\sqrt{\prod_{k=1}^{\ell}D_{d_1\,\ldots\,d_k}(x+\ell+1-k)}\cdot
  \frac{\text{W}[\phi_{d_1},\ldots,\phi_{d_{\ell}},\phi_n](x)}
  {\text{W}[\phi_{d_1},\ldots,\phi_{d_{\ell}}](x)},
  \label{phibarnD}\\
  \bar{B}(x)&=\sqrt{B(x+\ell)D(x+\ell+1)}\,
  \frac{\text{W}[\phi_{d_1},\ldots,\phi_{d_{\ell}}](x)}
  {\text{W}[\phi_{d_1},\ldots,\phi_{d_{\ell}}](x+1)}\,
  \frac{\text{W}[\phi_{d_1},\ldots,\phi_{d_{\ell}},\phi_{\mu}](x+1)}
  {\text{W}[\phi_{d_1},\ldots,\phi_{d_{\ell}},\phi_{\mu}](x)},
  \label{Bbar}\\
  \bar{D}(x)&=\sqrt{B(x-1)D(x)}\,
  \frac{\text{W}[\phi_{d_1},\ldots,\phi_{d_{\ell}}](x+1)}
  {\text{W}[\phi_{d_1},\ldots,\phi_{d_{\ell}}](x)}\,
  \frac{\text{W}[\phi_{d_1},\ldots,\phi_{d_{\ell}},\phi_{\mu}](x-1)}
  {\text{W}[\phi_{d_1},\ldots,\phi_{d_{\ell}},\phi_{\mu}](x)}.
  \label{Dbar}
\end{align}
{}From \eqref{Bsymp}--\eqref{Dsymp} we also have ($\ell\geq 0$)
\begin{align}
  \prod_{k=1}^{\ell}B_{d_1\,\ldots\,d_k}(x)&=
  \sqrt{\prod_{k=1}^{\ell}B(x+k-1)D(x+k)}\cdot
  \frac{\text{W}[\phi_{d_1},\ldots,\phi_{d_{\ell}}](x+1)}
  {\text{W}[\phi_{d_1},\ldots,\phi_{d_{\ell}}](x)},
  \label{Bsympl}\\
  \prod_{k=1}^{\ell}D_{d_1\,\ldots\,d_k}(x+\ell+1-k)&=
  \sqrt{\prod_{k=1}^{\ell}B(x+k-1)D(x+k)}\cdot
  \frac{\text{W}[\phi_{d_1},\ldots,\phi_{d_{\ell}}](x)}
  {\text{W}[\phi_{d_1},\ldots,\phi_{d_{\ell}}](x+1)}.
  \label{Dsympl}
\end{align}
Therefore the functions $\bar{B}(x)$, $\bar{D}(x)$ and $\bar{\phi}_n(x)$
are symmetric with respect to $d_1,\ldots,d_{\ell}$, and the final
Hamiltonian $\bar{\mathcal{H}}$ is independent of the order of $\{d_j\}$.

Let us state the discrete quantum mechanics version of Adler's theorem;
{\em If the set of deleted energy levels $\mathcal{D}=\{d_1,\ldots,d_{\ell}\}$
satisfy the condition \eqref{dellcond}, the modified Hamiltonian is given
by $\bar{\mathcal{H}}=\mathcal{H}_{d_1\,\ldots\,d_{\ell}}
=\bar{\mathcal{A}}^{\dagger}\bar{\mathcal{A}}+\mathcal{E}(\mu)$
with the potential functions given by \eqref{Bbar}--\eqref{Dbar} and
its eigenfunctions are given by\/} \eqref{phibarnB}--\eqref{phibarnD}.
The discrete QM version of Crum's theorem in \S\,\ref{sec:Crum} corresponds
to the choice $\{d_1,\ldots,d_{\ell}\}=\{0,1,\ldots,\ell-1\}$ and the new
groundstate is at the level $\mu=\ell$ and there is no vacant energy
level above that.
We have not yet proven the hermiticity of the resulting Hamiltonian
$\bar{\mathcal{H}}$ and the reality of the eigenfunctions $\bar{\phi}_n(x)$
categorically for the discrete quantum mechanics, even when the condition
\eqref{dellcond} is satisfied by the deleted levels.
This is due to the lack of the difference equation analogue of the
oscillation theorem.
It should be stressed that in most practical cases, in particular, in the
cases of polynomial eigenfunctions, the hermiticity of the Hamiltonian
$\bar{\mathcal{H}}$ is satisfied.

The orthogonality relation of the complete set of eigenfunctions is
\begin{equation}
  (\bar{\phi}_n,\bar{\phi}_m)\eqdef\sum_{x=0}^{\bar{x}_{\text{max}}}
  \bar{\phi}_n(x)\bar{\phi}_m(x)
  =\prod_{j=1}^{\ell}\bigl(\mathcal{E}(n)-\mathcal{E}({d_j})\bigr)\cdot
  \frac{1}{d_n^2}\,\delta_{nm}
  \quad (n,m\in\{0,1,\ldots,n_{\text{max}}\}\backslash\mathcal{D}).
  \label{adlortho}
\end{equation}
Note that the coefficient of $\delta_{nm}$ is positive for $\mathcal{D}$
satisfying \eqref{dellcond}.

\section{Simplifications}
\label{sec:simplifications}
\setcounter{equation}{0}

In the preceding sections, Crum's theorem and its modification are presented
in their most generic forms. These formulas are simplified substantially
when the system is {\em shape invariant\/} \cite{genden} and/or the
eigenfunctions consist of polynomials.

\subsection{Formulas for Crum's theorem}

\subsubsection{Shape invariance}
\label{sec:Crum:shapeinv}

Shape invariance is a well known sufficient condition for {\em exact
solvability\/} in ordinary quantum mechanics. The situation is exactly
the same in discrete quantum mechanics.
Shape invariance simply means that the $s$-th associated Hamiltonian
system has the same {\em form\/} as the original one with certain shifts
of the contained parameters.

The Hamiltonian may contain several parameters
$\bm{\lambda}=(\lambda_1,\lambda_2,$ $\ldots)$ and we indicate them
symbolically as
$\mathcal{H}=\mathcal{H}(\bm{\lambda})$,
$\mathcal{A}=\mathcal{A}(\bm{\lambda})$,
$\mathcal{E}(n)=\mathcal{E}(n;\bm{\lambda})$,
$\phi_n(x)=\phi_n(x;\bm{\lambda})$, etc.
Let us consider the case that the potential functions of the first
associated Hamiltonian $B^{[1]}(x)=B^{[1]}(x\,;\bm{\lambda})$ and
$D^{[1]}(x)=D^{[1]}(x\,;\bm{\lambda})$ have the same forms as the original
functions $B$ and $D$ with a different set of parameters and up to a
multiplicative positive constant $\kappa\in\mathbb{R}_{>0}$:
\begin{equation}
  B^{[1]}(x\,;\bm{\lambda})=\kappa B(x\,;\bm{\lambda}'),\quad
  D^{[1]}(x\,;\bm{\lambda})=\kappa D(x\,;\bm{\lambda}').
  \label{BDshape}
\end{equation}
Here the new set of parameters $\bm{\lambda}'$ is uniquely determined
by $\bm{\lambda}$. In concrete examples we can choose an appropriate set
of parameters such that $\bm{\lambda}'=\bm{\lambda}+\bm{\delta}$, with a
shift of parameters denoted by $\bm{\delta}$. In the following we assume this.
In fact, \eqref{BDshape} is equivalent to the definition of the shape
invariance by Odake and Sasaki \cite{os4,os12,os13},
\begin{equation}
  \mathcal{A}(\bm{\lambda})\mathcal{A}(\bm{\lambda})^{\dagger}
  =\kappa\mathcal{A}(\bm{\lambda}+\bm{\delta})^{\dagger}
  \mathcal{A}(\bm{\lambda}+\bm{\delta})
  +\mathcal{E}(1;\bm{\lambda}).
  \label{shapeinv1}
\end{equation}
It should be stressed that the above definition of shape invariance is
more stringent and thus more constraining than the original definition of
Gendenshtein \cite{genden}.
This condition is rewritten as
\begin{align}
  \sqrt{B(x+1;\bm{\lambda})D(x+1;\bm{\lambda})}
  &=\kappa\sqrt{B(x;\bm{\lambda}+\bm{\delta})D(x+1;\bm{\lambda}+\bm{\delta})},
  \label{shapeinv1cond1}\\
  B(x;\bm{\lambda})+D(x+1;\bm{\lambda})
  &=\kappa\bigl(B(x;\bm{\lambda}+\bm{\delta})+D(x;\bm{\lambda}+\bm{\delta})
  \bigr)+\mathcal{E}(1;\bm{\lambda}).
  \label{shapeinv1cond2}
\end{align}

The shape invariance condition \eqref{shapeinv1} combined with Crum's
theorem implies
\begin{align}
  &\mathcal{A}^{[s]}(\bm{\lambda})=\kappa^{\frac{s}{2}}
  \mathcal{A}(\bm{\lambda}+s\bm{\delta}),\quad
  \mathcal{A}^{[s]\dagger}(\bm{\lambda})=\kappa^{\frac{s}{2}}
  \mathcal{A}(\bm{\lambda}+s\bm{\delta})^{\dagger},\\
  &\mathcal{H}^{[s]}(\bm{\lambda})
  =\kappa^s\mathcal{H}(\bm{\lambda}+s\bm{\delta})
  +\mathcal{E}(s;\bm{\lambda}),
\end{align}
and the entire energy spectrum and the excited states wavefunctions are
expressed in terms of $\mathcal{E}(1;\bm{\lambda})$ and
$\phi_0(x\,;\bm{\lambda})$ as follows:
\begin{align}
  &\mathcal{E}(n;\bm{\lambda})=\sum_{s=0}^{n-1}
  \kappa^s\mathcal{E}(1;\bm{\lambda}+s\bm{\delta}),\\
  &\phi_n(x\,;\bm{\lambda})\propto
  \mathcal{A}(\bm{\lambda})^{\dagger}
  \mathcal{A}(\bm{\lambda}+\bm{\delta})^{\dagger}
  \mathcal{A}(\bm{\lambda}+2\bm{\delta})^{\dagger}
  \cdots
  \mathcal{A}(\bm{\lambda}+(n-1)\bm{\delta})^{\dagger}
  \phi_0(x\,;\bm{\lambda}+n\bm{\delta}).
  \label{Rodr}
\end{align}
Therefore the shape invariance is a sufficient condition for exact solvability.
We have also
\begin{align}
  &\mathcal{A}(\bm{\lambda})\phi_n(x;\bm{\lambda})
  =\frac{1}{\sqrt{B(0;\bm{\lambda})}}\,f_n(\bm{\lambda})
  \phi_{n-1}\bigl(x;\bm{\lambda}+\bm{\delta}\bigr),
  \label{Aphi=fphi}\\
  &\mathcal{A}(\bm{\lambda})^{\dagger}
  \phi_{n-1}\bigl(x;\bm{\lambda}+\bm{\delta}\bigr)
  =\sqrt{B(0;\bm{\lambda})}\,b_{n-1}(\bm{\lambda})\phi_n(x;\bm{\lambda}),
  \label{Adphi=bphi}
\end{align}
where $f_n(\bm{\lambda})$ and $b_{n-1}(\bm{\lambda})$ are the factors of
the energy eigenvalue,
$\mathcal{E}(n;\bm{\lambda})=f_n(\bm{\lambda})b_{n-1}(\bm{\lambda})$.
It is interesting to note that the polynomial eigenfunctions are not the
direct consequence of the shape invariance. When the eigenfunctions consist
of polynomials \eqref{phin=phi0Pn}, the above formula \eqref{Rodr} could be
called the Rodrigues formula for the polynomials.
Relations \eqref{Aphi=fphi}--\eqref{Adphi=bphi} would translate into the
{\em forward\/} and {\em backward shift relations\/} \cite{os12,os13} for
the polynomial eigenfunctions.

\subsubsection{Polynomial eigenfunctions}
\label{sec:Crum:poly}

Here we consider a generic Hamiltonian \eqref{genham}.
That is the shape invariance is not assumed.
Let us define a function $\eta(x)$ as a ratio of $\phi_1(x)$ and $\phi_0(x)$,
\begin{equation}
  \frac{\phi_1(x)}{\phi_0(x)}=a+b\,\eta(x),
  \label{etalinear}
\end{equation}
where $a$ and $b$ $(b\neq 0)$ are real constants.
Although $\eta(x)$ is not well defined without specifying $a$ and $b$,
this ambiguity (affine transformation of $\eta(x)$) does not affect
the following discussion. Then \eqref{B1} and \eqref{D1} imply
\begin{equation}
  B^{[1]}(x)=B(x+1)\,\frac{\eta(x+1)-\eta(x+2)}{\eta(x)-\eta(x+1)},\quad
  D^{[1]}(x)=D(x)\,\frac{\eta(x-1)-\eta(x)}{\eta(x)-\eta(x+1)}.
  \label{B1D1eta}
\end{equation}
Let us assume further that the $n$-th eigenfunction $\phi_n/\phi_0$ is
a degree $n$ polynomial in this $\eta(x)$ for all $n\ge2$:
\begin{equation}
  \phi_n(x)=\phi_0(x)P_n\bigl(\eta(x)\bigr),\quad
  P_n(y)=\sum_{k=0}^na_{n,k}\,y^k,\quad a_{n,n}\neq 0.
\end{equation}
Obviously $a_{0,0}=1$, $a=a_{1,0}$ and $b=a_{1,1}$.
The orthogonality of the eigenfunctions $\{\phi_n\}$ implies that
$\{P_n\bigl(\eta(x)\bigr)\}$ are orthogonal polynomials in $\eta(x)$
with respect to the weight function $\phi_0(x)^2$.
Then we can show the following:
\begin{align}
  &\frac{\phi^{[s]}_{s+1}(x)}{\phi^{[s]}_{s}(x)}
  =\frac{a_{s+1,s}}{a_{s,s}}+\frac{a_{s+1,s+1}}{a_{s,s}}\,\eta^{[s]}(x),
  \quad\eta^{[s]}(x)\eqdef\sum_{k=0}^s\eta(x+k),\\
  &B^{[s]}(x)=B^{[s-1]}(x+1)\,\frac{\eta^{[s-1]}(x+1)-\eta^{[s-1]}(x+2)}
  {\eta^{[s-1]}(x)-\eta^{[s-1]}(x+1)},\\
  &D^{[s]}(x)=D^{[s-1]}(x)\,\frac{\eta^{[s-1]}(x-1)-\eta^{[s-1]}(x)}
  {\eta^{[s-1]}(x)-\eta^{[s-1]}(x+1)},\\
  &\phi^{[s]}_n(x)=\phi^{[s]}_s(x)\check{\mathcal{P}}^{[s]}_n(x)\quad(n\geq s),
\end{align}
where $\check{\mathcal{P}}^{[s]}_n(x)$ is a symmetric polynomial of degree
$n-s$ in $\eta(x),\eta(x+1),\ldots,\eta(x+s)$ and satisfies the recurrence
relation
\begin{equation}
  \check{\mathcal{P}}^{[s]}_n(x)=\frac{a_{s-1,s-1}}{a_{s,s}}
  \frac{\check{\mathcal{P}}^{[s-1]}_n(x)-\check{\mathcal{P}}^{[s-1]}_n(x+1)}
  {\eta^{[s-1]}(x)-\eta^{[s-1]}(x+1)}.
  \label{ccPsn}
\end{equation}
Obviously $\check{\mathcal{P}}^{[0]}_n(x)=P_n(\eta(x))$ and
$\check{\mathcal{P}}^{[s]}_s(x)=1$.
{}From \eqref{ccPsn}, $\check{\mathcal{P}}^{[s]}_n(x)$ is explicitly
expressed as
\begin{equation}
  \check{\mathcal{P}}^{[s]}_n(x)=\frac{1}{a_{s,s}}\sum_{k=s}^na_{n,k}\!\!
  \sum_{\genfrac{}{}{0pt}{}{k_0,\ldots,k_s\geq 0}{\sum_{j=0}^sk_j=k-s}}
  \prod_{j=0}^s\eta(x+j)^{k_j}\quad(n\geq s\geq 0).
\end{equation}
For the shape invariant case, we have
$\check{\mathcal{P}}^{[s]}_n(x)\propto
P_{n-s}(\eta(x;\bm{\lambda}+s\bm{\delta});\bm{\lambda}+s\bm{\delta})$.

\subsubsection{Casorati Determinants}
\label{sec:det}

Here we prepare several formulas including various Casorati determinants
and the sinusoidal coordinates.

For any polynomial in $\eta$, $\{P_n(\eta)\}$ ($P_n(\eta)
=c_n\eta^n+(\text{lower degree terms})$), a set of variables $\{\eta_j\}$,
and a set of non-negative distinct integers $\{n_k\}$, it is easy to show
\begin{equation}
  \det\bigl(P_{n_k}(\eta_j)\bigr)_{1\leq j,k\leq m}
  =\!\!\prod_{1\leq j<k\leq m}(\eta_k-\eta_j)\cdot
  R(\eta_1,\ldots,\eta_m),
  \label{detPnk}
\end{equation}
where $R(\eta_1,\ldots,\eta_m)$ is a symmetric polynomial of degree
$\sum_{k=1}^mn_k-\frac12m(m-1)$ in $\eta_1,\ldots,\eta_m$.
Especially, for $(n_1,\ldots,n_m)=(0,1,\ldots,\ell)$,
the polynomial $R$ becomes a constant and we have
\begin{equation}
  \det\bigl(P_{k-1}(\eta_j)\bigr)_{1\leq j,k\leq\ell+1}
  =\prod_{k=0}^{\ell}c_k\,\cdot
  \!\!\!\prod_{1\leq j<k\leq\ell+1}(\eta_k-\eta_j).
  \label{det01..l}
\end{equation}

For all the sinusoidal coordinates $\eta(x)$ studied in \cite{os12,os14},
we have
\begin{align}
  \eta(x+\alpha)+\eta(x-\alpha)
  &=(\text{a polynomial of degree 1 in $\eta(x)$}),
  \label{eta+eta}\\
  \eta(x+\alpha)\eta(x-\alpha)
  &=(\text{a polynomial of degree 2 in $\eta(x)$}),
  \label{eta*eta}
\end{align}
with ${}^{\forall}\alpha\in\mathbb{R}$.
Hence a symmetric polynomial in $\eta(x),\eta(x+1),\ldots,\eta(x+\ell)$
becomes a polynomial in $\eta(x+\frac{\ell}{2})$.
Moreover the sinusoidal coordinate $\eta(x;\bm{\lambda})$ satisfies
\begin{equation}
  \eta(x+\alpha;\bm{\lambda})
  =(\text{a polynomial of degree 1 in 
  $\eta(x;\bm{\lambda}+2\alpha\bm{\delta})$}),
\end{equation}
with ${}^{\forall}\alpha\in\mathbb{R}$.
Therefore a symmetric polynomial in $\eta(x;\bm{\lambda}),
\eta(x+1;\bm{\lambda}),\ldots,\eta(x+\ell;\bm{\lambda})$ becomes a
polynomial in $\eta(x;\bm{\lambda}+\ell\bm{\delta})$.
The sinusoidal coordinate $\eta(x;\bm{\lambda})$ satisfies
\begin{equation}
  \frac{\eta(x+\alpha;\bm{\lambda})-\eta(x;\bm{\lambda})}
  {\eta(\alpha;\bm{\lambda})}
  =\varphi(x;\bm{\lambda}+(\alpha-1)\bm{\delta}).
  \label{eta-eta}
\end{equation}
For the definition of the auxiliary function $\varphi(x;\bm{\lambda})$,
see eqs.\,(4.12) and (4.23) in \cite{os12}.
Let us define the $\ell$-th auxiliary function
$\varphi_{\ell}(x;\bm{\lambda})$ as
\begin{equation}
  \varphi_{\ell}(x;\bm{\lambda})\eqdef
  \!\!\prod_{0\leq j<k\leq\ell-1}
  \frac{\eta(x+k;\bm{\lambda})-\eta(x+j;\bm{\lambda})}{\eta(k-j;\bm{\lambda})}
  =\!\!\prod_{0\leq j<k\leq\ell-1}\varphi(x+j;\bm{\lambda}+(k-j-1)\bm{\delta}).
\end{equation}
Then, for any polynomials of degree $n$ in $\eta(x;\bm{\lambda})$,
$\check{P}_n(x;\bm{\lambda})\eqdef P_n(\eta(x;\bm{\lambda});\bm{\lambda})$,
we have
\begin{equation}
  \frac{\text{W}[\check{P}_{n_1},\check{P}_{n_2},\ldots,\check{P}_{n_m}]
  (x;\bm{\lambda})}{\varphi_m(x;\bm{\lambda})}=\genfrac{(}{)}{0pt}{}
  {\text{a polynomial of degree $\sum_{k=1}^mn_k-\frac12m(m-1)$}}
  {\text{in $\eta(x;\bm{\lambda}+(m-1)\bm{\delta})$}},
  \label{ppolydef}
\end{equation}
and \eqref{det01..l} implies
\begin{equation}
  \text{W}[\check{P}_0,\check{P}_1,\ldots,\check{P}_{\ell}](x;\bm{\lambda})
  =\prod_{k=0}^{\ell}c_k(\bm{\lambda})\,\cdot
  \prod_{k=1}^{\ell}\prod_{j=1}^k\eta(j;\bm{\lambda})\cdot
  \varphi_{\ell+1}(x;\bm{\lambda}).
  \label{W12..l0}
\end{equation}
Note that, for the orthogonal polynomials studied in \cite{os12},
we have \eqref{WP}.

\subsection{Formulas for modified Crum's theorem}
\label{sec:Adler:poly}

In this subsection we consider the simplification of the formulas in
\S\,\ref{sec:Adler} when the eigenfunctions consist of the orthogonal
polynomials $\{P_n\}$:
\begin{equation}
  \phi_n(x)=\phi_0(x)P_n(\eta(x)),\quad \check{P}_n(x)\eqdef P_n(\eta(x)),
  \label{phin=phi0Pn,cP}
\end{equation}
where $P_n(\eta)$ is a polynomial of degree $n$ in $\eta$, satisfying
the boundary condition $\eta(0)=0$ \eqref{Pnorm}.

First we recall that the ground state $\phi_0(x)$ is annihilated by
$\mathcal{A}$,
\begin{equation}
  \sqrt{B(x)}\,\phi_0(x)=\sqrt{D(x+1)}\,\phi_0(x+1).
\end{equation}
By using this and \eqref{Wformula1}, eqs.\,\eqref{Bsympl}--\eqref{Dsympl}
become
\begin{align}
  \prod_{k=1}^{\ell}B_{d_1\,\ldots\,d_k}(x)&=
  \prod_{k=1}^{\ell}B(x+k-1)\cdot
  \frac{\text{W}[\check{P}_{d_1},\ldots,\check{P}_{d_{\ell}}](x+1)}
  {\text{W}[\check{P}_{d_1},\ldots,\check{P}_{d_{\ell}}](x)},
  \label{Bsymplp}\\
  \prod_{k=1}^{\ell}D_{d_1\,\ldots\,d_k}(x+\ell+1-k)&=
  \prod_{k=1}^{\ell}D(x+k)\cdot
  \frac{\text{W}[\check{P}_{d_1},\ldots,\check{P}_{d_{\ell}}](x)}
  {\text{W}[\check{P}_{d_1},\ldots,\check{P}_{d_{\ell}}](x+1)}.
  \label{Dsymplp}
\end{align}
{}From these and \eqref{phibarnB}--\eqref{Dbar} we obtain
\begin{align}
  \bar{\phi}_n(x)&=
  (-1)^{\ell}\phi_0(x)\sqrt{\prod_{k=1}^{\ell}B(x+k-1)}\cdot
  \frac{\text{W}[\check{P}_{d_1},\ldots,\check{P}_{d_{\ell}},\check{P}_n](x)}
  {\sqrt{\text{W}[\check{P}_{d_1},\ldots,\check{P}_{d_{\ell}}](x)
  \text{W}[\check{P}_{d_1},\ldots,\check{P}_{d_{\ell}}](x+1)}}
  \label{phibarnBp}\\
  &=(-1)^{\ell}\phi_0(x+\ell)\sqrt{\prod_{k=1}^{\ell}D(x+k)}\cdot
  \frac{\text{W}[\check{P}_{d_1},\ldots,\check{P}_{d_{\ell}},\check{P}_n](x)}
  {\sqrt{\text{W}[\check{P}_{d_1},\ldots,\check{P}_{d_{\ell}}](x)
  \text{W}[\check{P}_{d_1},\ldots,\check{P}_{d_{\ell}}](x+1)}},\!\!
  \label{phibarnDp}\\
  \bar{B}(x)&=B(x+\ell)\,
  \frac{\text{W}[\check{P}_{d_1},\ldots,\check{P}_{d_{\ell}}](x)}
  {\text{W}[\check{P}_{d_1},\ldots,\check{P}_{d_{\ell}}](x+1)}\,
  \frac{\text{W}[\check{P}_{d_1},\ldots,\check{P}_{d_{\ell}},
  \check{P}_{\mu}](x+1)}
  {\text{W}[\check{P}_{d_1},\ldots,\check{P}_{d_{\ell}},\check{P}_{\mu}](x)},
  \label{Bbarp}\\
  \bar{D}(x)&=D(x)\,
  \frac{\text{W}[\check{P}_{d_1},\ldots,\check{P}_{d_{\ell}}](x+1)}
  {\text{W}[\check{P}_{d_1},\ldots,\check{P}_{d_{\ell}}](x)}\,
  \frac{\text{W}[\check{P}_{d_1},\ldots,\check{P}_{d_{\ell}},
  \check{P}_{\mu}](x-1)}
  {\text{W}[\check{P}_{d_1},\ldots,\check{P}_{d_{\ell}},\check{P}_{\mu}](x)}.
  \label{Dbarp}
\end{align}

The eigenfunction $\bar{\phi}_n(x)$ of this $\mathcal{D}$-deleted system
has the following structure,
\begin{equation}
  \bar{\phi}_n(x)=\bar{\phi}_{\mu}(x)\times
  \frac{\text{W}[\check{P}_{d_1},\ldots,\check{P}_{d_{\ell}},\check{P}_n](x)}
  {\text{W}[\check{P}_{d_1},\ldots,\check{P}_{d_{\ell}},\check{P}_{\mu}](x)}.
  \label{defpolys}
\end{equation}
In the following we assume that $\eta(x)$ satisfies
\eqref{eta+eta}--\eqref{eta-eta}.
Then the above formulas \eqref{phibarnBp}--\eqref{Dbarp} are simplified
thanks to \eqref{ppolydef}:
\begin{align}
  \bar{\phi}_n(x;\bm{\lambda})&=\bar{\phi}_{\mu}(x;\bm{\lambda})\times
  \frac{\mathcal{P}_n(\eta_{\ell}(x))}
  {\mathcal{P}_{\mu}(\eta_{\ell}(x))},\qquad
  \eta_{\ell}(x)\eqdef\eta(x;\bm{\lambda}+\ell\bm{\delta}),
  \label{defphibarn}\\
  \bar{B}(x)&=B(x+\ell)\,\frac{\mathcal{P}(\eta_{\ell-1}(x))}
  {\mathcal{P}(\eta_{\ell-1}(x+1))}
  \frac{\mathcal{P}_{\mu}(\eta_{\ell}(x+1))}{\mathcal{P}_{\mu}(\eta_{\ell}(x))}
  \frac{\varphi_{\ell}(x;\bm{\lambda})}{\varphi_{\ell}(x+1;\bm{\lambda})}
  \frac{\varphi_{\ell+1}(x+1;\bm{\lambda})}{\varphi_{\ell+1}(x;\bm{\lambda})},
  \label{Bbarppoly}\\
  \bar{D}(x)&=D(x)\,\frac{\mathcal{P}(\eta_{\ell-1}(x+1))}
  {\mathcal{P}(\eta_{\ell-1}(x))}
  \frac{\mathcal{P}_{\mu}(\eta_{\ell}(x-1))}{\mathcal{P}_{\mu}(\eta_{\ell}(x))}
  \frac{\varphi_{\ell}(x+1;\bm{\lambda})}{\varphi_{\ell}(x;\bm{\lambda})}
  \frac{\varphi_{\ell+1}(x-1;\bm{\lambda})}{\varphi_{\ell+1}(x;\bm{\lambda})},
  \label{Dbarppoly}
\end{align}
where the deformed polynomials $\mathcal{P}(\eta_{\ell-1})$,
$\mathcal{P}_n(\eta_{\ell})$ and $\mathcal{P}_\mu(\eta_{\ell})$ are defined
as (we suppress writing $\bm{\lambda}$-dependence explicitly)
\begin{align}
  \mathcal{P}(\eta_{\ell-1}(x))&\eqdef
  \varphi_{\ell}(x;\bm{\lambda})^{-1}\,
  \text{W}[\check{P}_{d_1},\ldots,\check{P}_{d_{\ell}}](x),
  \label{mPdef}\\
  \mathcal{P}_{\mu}(\eta_{\ell}(x))&\eqdef
  \varphi_{\ell+1}(x;\bm{\lambda})^{-1}\,
  \text{W}[\check{P}_{d_1},\ldots,\check{P}_{d_{\ell}},\check{P}_\mu](x),
  \label{mPmudef}\\
  \mathcal{P}_n(\eta_{\ell}(x))&\eqdef
  \varphi_{\ell+1}(x;\bm{\lambda})^{-1}
  \text{W}[\check{P}_{d_1},\ldots,\check{P}_{d_{\ell}},\check{P}_n](x),
  \label{mPndef}
\end{align}
and they are of degree $|\mathcal{D}|-\frac12\ell(\ell-1)$,
$|\mathcal{D}|+\mu-\frac12\ell(\ell+1)$ and
$|\mathcal{D}|+n-\frac12\ell(\ell+1)$
in $\eta_{\ell-1}$, $\eta_{\ell}$ and $\eta_{\ell}$, respectively.
Here $|\mathcal{D}|$ is defined by $|\mathcal{D}|\eqdef\sum_{j=1}^\ell d_j$.
By construction the deformed polynomials $\mathcal{P}(\eta_{\ell-1}(x))$
and $\mathcal{P}_\mu(\eta_{\ell}(x))$, which appears in the denominator
of \eqref{defphibarn} and in \eqref{Bbarppoly}-\eqref{Dbarppoly}, are
positive definite for $x=0,1,\ldots,\bar{x}_{\text{max}}$.

The ratio of the polynomials
$\mathcal{P}_n(\eta_{\ell}(x))/\mathcal{P}_\mu(\eta_{\ell}(x))$
are eigenfunctions of the similarity transformed Hamiltonian
(a second order difference operator)
$\widetilde{\bar{\mathcal{H}}}$:
\begin{align}
  &\widetilde{\bar{\mathcal{H}}}\eqdef
  \bar{\phi}_{\mu}(x;\bm{\lambda})^{-1}\circ\bar{\mathcal{H}}\circ
  \bar{\phi}_{\mu}(x;\bm{\lambda})
  =\bar{B}(x)(1-e^{\partial})+\bar{D}(x)(1-e^{-\partial}),\\
  &\widetilde{\bar{\mathcal{H}}}\,
  \frac{\mathcal{P}_n(\eta_{\ell}(x))}{\mathcal{P}_\mu(\eta_{\ell}(x))}
  =\mathcal{E}(n)
  \frac{\mathcal{P}_n(\eta_{\ell}(x))}{\mathcal{P}_\mu(\eta_{\ell}(x))}
  \quad(n\in\{0,1,\ldots,n_{\text{max}}\}\backslash{\mathcal{D}}).
  \label{defSch}
\end{align}
The orthogonality relations of $\{\bar{\phi}_n\}$ \eqref{adlortho} are now
rewritten as the orthogonality relations of the polynomials
$\{\mathcal{P}_n(\eta_{\ell}(x))\}$,
\begin{equation}
  \sum_{x=0}^{\bar{x}_{\text{max}}}
  \bar{\psi}(x;\bm{\lambda})^2\mathcal{P}_n(\eta_{\ell}(x))
  \mathcal{P}_m(\eta_{\ell}(x))
  =\prod_{j=1}^{\ell}\bigl(\mathcal{E}(n)-\mathcal{E}({d_j})\bigr)\cdot
  \frac{1}{d_n^2}\,\delta_{nm}
  \quad(n,m\in\{0,1,\ldots,n_{\text{max}}\}\backslash\mathcal{D}),
  \label{orthocPcP}
\end{equation}
in which the new weightfunction is defined by
\begin{equation}
  \bar{\psi}(x;\bm{\lambda})\eqdef
  \frac{\bar{\phi}_\mu(x;\bm{\lambda})}{\mathcal{P}_\mu(\eta_{\ell}(x))}.
  \label{defpsi}
\end{equation}

We have demonstrated the following:
When the modified Crum's theorem is applied to the shape invariant
(exactly solvable) Hamiltonian systems corresponding to the ($q$)-Askey
scheme of hypergeometric orthogonal polynomials $P_n(\eta(x))$ \cite{os12},
the resulting system is no longer shape invariant but it is exactly
solvable since all the eigenvalues and the corresponding eigenfunctions
are explicitly given. The eigenfunctions
consist of deformed polynomials $\mathcal{P}_{n}(\eta_{\ell}(x))$
\eqref{mPndef}. They inherit the sinusoidal coordinate of the original
system and they are orthogonal
with respect to the weight function given as the square of the deformed
groundstate eigenfunction $\bar{\psi}(x;\bm{\lambda})^2$ \eqref{defpsi}.
These deformed orthogonal polynomials $\mathcal{P}_{n}(\eta_{\ell}(x))$,
although forming a complete set of eigenpolynomials,
have vacancies in the degrees corresponding
to the deleted levels $\mathcal{D}=\{d_1,d_2,\ldots,d_{\ell}\}$.
Therefore they {\em do not satisfy three term recurrence relations\/} and
they are not called orthogonal polynomials in the strict sense.
When another (modified) Crum's theorem is applied to such a system, 
most of the simplification formulas (due to polynomial eigenfunctions)
in this section are still valid with some adjustments.

\section{Dual Christoffel Transformations }
\label{sec:dualChris}
\setcounter{equation}{0}

In this section we will show that the simple case ({\em i.e.\/}\ $\mu=0$)
of the modified Crum's theorem applied to
an orthogonal polynomial $P_n(\eta(x))$ provides the well known Christoffel
transformation \cite{chris,askey} for the corresponding dual orthogonal
polynomial $Q_x(\mathcal{E}(n))$ \eqref{duality}. Therefore we will call
it the {\em dual Christoffel transformation\/}.
Corresponding to the deletion of the {\em levels} or {\em degrees}
$n\in\mathcal{D}=\{d_1,\ldots,d_{\ell}\}$ of the polynomials $P_n(\eta(x))$,
the {\em positions} $n\in\mathcal{D}=\{d_1,\ldots,d_{\ell}\}$ are deleted
from the dual polynomial $Q_x(\mathcal{E}(n))$.
Here we assume that the set of deleted levels does not contain 0,
the original ground state, {\em i.e.} $0\notin\mathcal{D}$.
This also means that $\ell$ is {\em even\/} and the modified groundstate
has also the label 0, $\mu=0$.
The dual Christoffel transformation needs the framework of the discrete
quantum mechanics and it cannot be applied to the general orthogonal
polynomials.

The deformed dual polynomials  $\mathcal{Q}_{x}(\mathcal{E})$ are defined
by the three term recurrence relations in terms of $\bar{B}(x)$
\eqref{Bbarp} and $\bar{D}(x)$ \eqref{Dbarp}:
\begin{align}
  &\bar{B}(x)\bigl(\mathcal{Q}_{x}(\mathcal{E})
  -\mathcal{Q}_{x+1}(\mathcal{E})\bigr)
  +\bar{D}(x)\bigl(\mathcal{Q}_{x}(\mathcal{E})
  -\mathcal{Q}_{x-1}(\mathcal{E})\bigr)
  =\mathcal{E}\mathcal{Q}_{x}(\mathcal{E})\quad
  (x=0,1,\ldots,\bar{x}_{\text{max}}),\n
  &\mathcal{Q}_{0}(\mathcal{E})=1,\quad
  \mathcal{Q}_{-1}(\mathcal{E})=0.
  \label{def3term}
\end{align}
One simple consequence of the above recurrence relation is the universal
normalisation
\begin{equation}
  \mathcal{Q}_{x}(0)=1\quad(x=0,1,\ldots,\bar{x}_{\text{max}}).
\end{equation}
Assume that $\eta(x)$ satisfies \eqref{eta+eta}--\eqref{eta-eta}.
At the eigenvalues $\mathcal{E}=\mathcal{E}(n)$, the Schr\"odinger equation
\eqref{defSch} for the ratio of the deformed polynomials
$\mathcal{P}_n(\eta_{\ell}(x))/\mathcal{P}_0(\eta_{\ell}(x))$ is identical
with the above three term recurrence relation and the {\em duality relation}
holds:
\begin{equation}
  \frac{\mathcal{P}_n(\eta_{\ell}(x))}{\mathcal{P}_0(\eta_{\ell}(x))}
  =p_n\mathcal{Q}_x(\mathcal{E}(n)),\quad
  x\in [0,\bar{x}_{\text{max}}],\quad
  n\in\{0,1,\ldots,n_{\text{max}}\}\backslash\mathcal{D},
  \label{defdual}
\end{equation}
in which
\begin{equation}
  p_n\eqdef\frac{\mathcal{P}_{n}(0)}{\mathcal{P}_{0}(0)}
  =(-1)^\ell\prod_{j=1}^{\ell}\frac{\mathcal{E}(n)-\mathcal{E}({d_j})}
  {\mathcal{E}({d_j})},\quad
   n\in\{0,1,\ldots,n_{\text{max}}\}\backslash\mathcal{D},
  \label{defpn}
\end{equation}
by the normalisation condition. We have verified the second equality
for all the examples in \cite{os12}, see \eqref{WP}--\eqref{WP01}.
Thus $\mathcal{Q}_{x}(\mathcal{E})$ are genuine orthogonal polynomials,
satisfying the orthogonality relation dual to \eqref{orthocPcP}:
\begin{align}
  \sum_{n\in\{0,1,\ldots,n_{\text{max}}\}\backslash\mathcal{D}}
  &d_n^2\prod_{j=1}^{\ell}\bigl(\mathcal{E}(n)-\mathcal{E}({d_j})\bigr)\cdot
  \mathcal{Q}_x(\mathcal{E}(n))\mathcal{Q}_y(\mathcal{E}(n))\n
  &=\frac{\prod_{j=1}^{\ell}\mathcal{E}(d_j)^2}{\bar{\phi}_0(x)^2}
  \,\delta_{xy}\quad(x,y=0,1,\ldots,\bar{x}_{\text{max}}).
  \label{deforthoQ}
\end{align}
Note that $\prod_{j=1}^{\ell}\bigl(\mathcal{E}(n)-\mathcal{E}({d_j})\bigr)>0$
due to \eqref{dellcond}.
It is obvious that the values of $\mathcal{Q}_{x}(\mathcal{E}(n))$ at
`{\em positions}' $n\in\mathcal{D}=\{d_1,d_2,\ldots,d_{\ell}\}$ do not
enter the orthogonality relation or the Schr\"odinger equation \eqref{defSch}.
In other words, the values of $\mathcal{Q}_{x}(\mathcal{E}(n))$ at
$n\in\mathcal{D}$ are not defined but the degree $x$ runs from 0 to
$\bar{x}_{\text{max}}$ without any hole.
This transformation of the dual orthogonal polynomials,
${Q}_{x}(\mathcal{E}(n))\to\mathcal{Q}_{x}(\mathcal{E}(n))$ is the well
known {\em Christoffel transformation\/} \cite{chris,askey,geroni,yermozhed,
spirivinetzhed}.
The transformation of the orthogonality weight function is read from
\eqref{deforthoQ}:
\begin{equation}
  d_n^2\to
  d_n^2\prod_{j=1}^{\ell}\bigl(\mathcal{E}(n)-\mathcal{E}({d_j})\bigr).
\end{equation}
The inverse of the Christoffel transformation is called the Geronimus
transformation \cite{geroni}, which adds new `positions'. Its effects can be
practically incorporated by the dual Christoffel transformations with
redefinion of the parameters.

The dual Christoffel transformation has the merits that the formulas
determining $\bar{B}(x)$ and $\bar{D}(x)$ \eqref{Bbarp} and \eqref{Dbarp}
are universal, concise and algorithmic compared with the original Christoffel
transformation, which should be performed specifically for each case,
the polynomials and the set of deletion $\mathcal{D}$.
While the dual Christoffel transformation requires the framework of the
discrete quantum mechanics, the original Christoffel transformation is
defined for general orthogonal polynomials, {\em i.\,e.,} those having
the three term recurrence relations only without the difference equation,
nor the sinusoidal coordinate.
The elementary Christoffel transformation is defined by
\begin{equation}
  \widetilde{P}_n(x)\eqdef\frac{P_{n+1}(x)-A_nP_n(x)}{x-a},\quad
  A_n\eqdef\frac{P_{n+1}(a)}{P_n(a)},
\end{equation}
for a parameter $a\in\mathbb{R}$, which is not constrained by the sinusoidal
coordinate. Corresponding to the $\ell$-deletion above, one considers
multiple ($\ell$ times) applications of the elementary Christoffel
transformations. The normalisation weight function for the polynomials
$\{P_n(x)\}$ is, in general, not known explicitly and it usually contains
continuous measure. Thus determination of the positivity of the resulting
weight function generically involves a moment problem.
For the polynomials belonging to the discrete quantum mechanics
\cite{os12,koeswart}, explicit forms of multiple Christoffel
transformations are known for some specific cases. For example, see
\cite{yermozhed,spirivinetzhed} for the Racah polynomials and others.

\section{Summary and Comments}
\label{summary}
\setcounter{equation}{0}

Crum's theorem and its modification \`a la Adler are formulated for the
discrete quantum mechanics with real shifts.
They are slightly more complicated than those in the ordinary quantum
mechanics or in the discrete quantum mechanics with pure imaginary shifts,
partly because the size of the Hamiltonian matrix, or the range of the
$x$ variable ($x_{\text{max}}$), is reduced by one in each deletion.
Another source of complications is the fact that in the generic cases
the sinusoidal coordinate $\eta(x)$ depends on the parameters $\bm{\lambda}$,
which changes at each deletion. In the discrete quantum mechanics with
real shifts, the modified Crum's theorem generates the Christoffel
transformation of the dual orthogonal polynomials.
Very special and simple examples, in which all the excited states from
the first to the $\ell$-th are deleted (see Fig.\,3), are presented
explicitly in Appendix for more than two dozens of orthogonal polynomials
discussed in \cite{os12}.

One of the motivations of the present research is the connection with
the {\em infinitely many exceptional orthogonal polynomials}
\cite{gomez,quesne,os16,gomez2,os17}.
As explained in some detail in \S\,4 of \cite{gos}, the insight obtained
from the explicit examples of the application of modified Crum's theorem
was instrumental for the discovery of the infinitely many exceptional
Laguerre and Jacobi polynomials in the ordinary quantum mechanics
\cite{os16} and the exceptional Wilson and Askey-Wilson polynomials
in the discrete quantum mechanics with the pure imaginary shifts \cite{os17}.
The explicit examples in the present paper are also very helpful for
the discovery of the (infinitely) many {\em exceptional orthogonal
polynomials of a discrete variable}, for example, the exceptional
$q$-Racah polynomials, etc. They will be derived and discussed in detail
elsewhere \cite{os23}.

\section*{Acknowledgements}

R.\,S. is supported in part by Grant-in-Aid for Scientific Research
from the Ministry of Education, Culture, Sports, Science and Technology
(MEXT), No.19540179.

\bigskip
\appendix
\renewcommand{\theequation}{\Alph{section}.\arabic{equation}}
\section{Special Examples}
\label{sec:Ex}
\setcounter{equation}{0}

In Appendix we present very special and simple examples of an application
of Adler's theorem, in which the eigenstates
$\phi_1,\phi_2,\ldots,\phi_{\ell}$ are deleted. Similar examples of the
application of Adler's theorem in discrete quantum mechanics with pure
imaginary shifts were given in Appendix B of \cite{gos}.
In the ordinary quantum mechanics, some simple examples of the same sort
were demonstrated in Appendix A of \cite{gos} and \cite{dubov}.

In this case, $\mathcal{D}=\{d_1,d_2,\ldots,d_{\ell}\}=\{1,2,\ldots,\ell\}$,
that is, the modified groundstate level is the same as that of the original
theory $\mu=0$.
This means $|\mathcal{D}|=\frac12\ell(\ell+1)$ and the polynomial
$\mathcal{P}_{\mu}$ \eqref{mPmudef} is a constant.
The degrees of the deformed polynomial $\mathcal{P}$ \eqref{mPdef} and
$\mathcal{P}_n$ \eqref{mPndef} are $\ell$ and $\ell+n$, and they are
proportional to the deforming polynomial $\xi_{\ell}$
and the deformed polynomial $P_{\ell,n}$ \eqref{xilplndef}, respectively.
The duality \eqref{defdual} is now a relation between two polynomials
$\mathcal{P}_n(\eta_{\ell})$ and $\mathcal{Q}_x(\mathcal{E})$.

The main results are the unified expressions of the functions $B_{\ell}(x)$
\eqref{Blform} and $D_{\ell}(x)$ \eqref{Dlform}, which specify the Hamiltonian
$\mathcal{H}_{\ell}$ together with the eigenfunctions $\phi_{\ell,0}$
\eqref{phil0form} and $\phi_{\ell,n}(x)$ \eqref{philnform}.
The unified expressions of the deformed eigenpolynomials
$\check{P}_{\ell,n}(x)$ are given in \eqref{plnform}.
An important ingredient specific to each type of polynomials is the
{\em deforming polynomial\/} $\check{\xi}_{\ell}(x)$, which is listed
in \ref{sec:xil}.
The situation is illustrated in Fig.\,3, which should be compared with
Fig.\,2a depicting the generic case discussed in section \ref{sec:Adler}.

\begin{figure}[htbp]
\begin{center}
  \includegraphics{adlerfig3new.epsi}\\
  Figure\,3: Special case
  \label{default}
\end{center}
\end{figure}

The black circles denote the energy levels, whereas the white circles
denote {\em deleted\/} energy levels.
We write $\bar{\mathcal{H}}=\mathcal{H}_{12\,\ldots\,\ell}$,
$\bar{\phi}_n=\phi_{12\,\ldots\,\ell\,n}$,
$\bar{\mathcal{A}}=\mathcal{A}_{12\,\ldots\,\ell}$,
$\bar{B}=B_{12\,\ldots\,\ell}$,
$\bar{D}=D_{12\,\ldots\,\ell}$ etc.\ as $\mathcal{H}_{\ell}$,
$\phi_{\ell,n}$, $\mathcal{A}_{\ell}$, $B_{\ell}$, $D_{\ell}$ etc.
This Hamiltonian $\mathcal{H}_{\ell}
=\mathcal{A}_{\ell}^\dagger\mathcal{A}_{\ell}$ is hermitian for even
$\ell$ but may be non-hermitian for odd $\ell$.
Since algebraic formulas such as the Casoratians are valid for even and
odd $\ell$, we present various formulas without restricting to the even
$\ell$. The original systems are shape invariant but the
$(\phi_1,\ldots,\phi_{\ell})$-deleted systems $\mathcal{H}_{\ell}$ are not.
The rightmost vertical line in Fig.\,3 corresponds to the Hamiltonian
system $\mathcal{H}_{\ell}'=\mathcal{A}_{\ell}\mathcal{A}_{\ell}^\dagger
=\mathcal{H}^{[\ell+1]}$,
which is shape invariant and it is obtained from $\mathcal{H}_{\ell}$ by
one more step of Crum's method.

We apply Adler's theorem to the shape invariant, therefore solvable,
systems whose eigenfunctions are described by the orthogonal polynomials
studied in \cite{os12}; {\em i.e.}\ ($q$)-Racah, ($q$)-(dual)-Hahn, etc.
Hereafter we display the parameter dependence explicitly by $\bm{\lambda}$,
which represents the set of the parameters, and we follow the notation
of \cite{os12}. This is to emphasise that the basic structure of Crum's
theorem and its modification discussed in the main sections are well-founded
for the most generic systems.

\subsection{The original systems}
\label{sec:orig}

Here we summarise various properties of the original Hamiltonian systems
studied in \cite{os12} to be compared with the specially modified systems
to be presented in \ref{sec:12..l_delete}.
All the examples are shape invariant and exactly solvable.
Let us start with the Hamiltonians, Schr\"{o}dinger equations and
eigenfunctions ($x=0,1,\ldots,x_{\text{max}}$,
$x_{\text{max}}=\text{$N$ or $\infty$}$,
$n_{\text{max}}=\text{$N$ or $\infty$}$):
\begin{align}
  &\mathcal{A}(\bm{\lambda})\eqdef
  \sqrt{B(x;\bm{\lambda})}
  -e^{\partial}\sqrt{D(x;\bm{\lambda})},\quad
  \mathcal{A}(\bm{\lambda})^{\dagger}=
  \sqrt{B(x;\bm{\lambda})}
  -\sqrt{D(x;\bm{\lambda})}\,e^{-\partial},\\
  &\mathcal{H}(\bm{\lambda})\eqdef
  \mathcal{A}(\bm{\lambda})^{\dagger}\mathcal{A}(\bm{\lambda}),\\
  &\mathcal{H}(\bm{\lambda})\phi_n(x;\bm{\lambda})
  =\mathcal{E}(n;\bm{\lambda})\phi_n(x;\bm{\lambda})\quad
  (n=0,1,\ldots,n_{\text{max}}),\\
  &\phi_n(x;\bm{\lambda})=\phi_0(x;\bm{\lambda})
  P_n(\eta(x;\bm{\lambda});\bm{\lambda})\quad
  \bigl(P_n(0;\bm{\lambda})=1,\ \ \check{P}_n(x;\bm{\lambda})\eqdef
  P_n(\eta(x;\bm{\lambda});\bm{\lambda})\bigr).
\end{align}
The explicit forms of the set of parameters $\bm{\lambda}$, the potential
functions $B(x;\bm{\lambda})$ and $D(x;\bm{\lambda})$,
the energy eigenvalues $\mathcal{E}(n;\bm{\lambda})$,
the sinusoidal coordinate $\eta(x;\bm{\lambda})$, the ground state
wavefunctions $\phi_0(x;\bm{\lambda})$ and the eigenpolynomials
$P_n(\eta(x;\bm{\lambda});\bm{\lambda})$ are given in \cite{os12}.
The groundstate wavefunction $\phi_0(x;\bm{\lambda})$ is annihilated by
$\mathcal{A}(\bm{\lambda})$,
$\mathcal{A}(\bm{\lambda})\phi_0(x;\bm{\lambda})=0$, and given by
\begin{equation}
  \phi_0(x;\bm{\lambda})=\sqrt{\prod_{y=0}^{x-1}
  \frac{B(y;\bm{\lambda})}{D(y+1;\bm{\lambda})}}.
\end{equation}

The systems are shape invariant,
\begin{equation}
  \mathcal{A}(\bm{\lambda})\mathcal{A}(\bm{\lambda})^{\dagger}
  =\kappa\mathcal{A}(\bm{\lambda+\bm{\delta}})^{\dagger}
  \mathcal{A}(\bm{\lambda}+\bm{\delta})
  +\mathcal{E}(1;\bm{\lambda}),
\end{equation}
where the explicit forms of $\bm{\delta}$ and $\kappa$ are given in \cite{os12}.
The action of $\mathcal{A}(\bm{\lambda})$ and
$\mathcal{A}(\bm{\lambda})^{\dagger}$ on the eigenfunctions is
\begin{align}
  &\mathcal{A}(\bm{\lambda})\phi_n(x;\bm{\lambda})
  =\frac{1}{\sqrt{B(0;\bm{\lambda})}}\,f_n(\bm{\lambda})
  \phi_{n-1}\bigl(x;\bm{\lambda}+\bm{\delta}\bigr),\\
  &\mathcal{A}(\bm{\lambda})^{\dagger}
  \phi_{n-1}\bigl(x;\bm{\lambda}+\bm{\delta}\bigr)
  =\sqrt{B(0;\bm{\lambda})}\,b_{n-1}(\bm{\lambda})\phi_n(x;\bm{\lambda}),
\end{align}
where $f_n(\bm{\lambda})$ and $b_{n-1}(\bm{\lambda})$ are the factors of
the energy eigenvalue,
$\mathcal{E}(n;\bm{\lambda})=f_n(\bm{\lambda})b_{n-1}(\bm{\lambda})$,
and their explicit forms are given in \cite{os12}.
The forward and backward shift operators $\mathcal{F}(\bm{\lambda})$ and
$\mathcal{B}(\bm{\lambda})$ are defined by
\begin{align}
  \mathcal{F}(\bm{\lambda})&\eqdef
  \sqrt{B(0;\bm{\lambda})}\,\phi_0(x;\bm{\lambda}+\bm{\delta})^{-1}\circ
  \mathcal{A}(\bm{\lambda})\circ\phi_0(x;\bm{\lambda})\n
  &=B(0;\bm{\lambda})\varphi(x;\bm{\lambda})^{-1}(1-e^{\partial}),\\
  \mathcal{B}(\bm{\lambda})&\eqdef
  \frac{1}{\sqrt{B(0;\bm{\lambda})}}\,\phi_0(x;\bm{\lambda})^{-1}\circ
  \mathcal{A}(\bm{\lambda})^{\dagger}
  \circ\phi_0(x;\bm{\lambda}+\bm{\delta})\n
  &=\frac{1}{B(0;\bm{\lambda})}\bigl(B(x;\bm{\lambda})
  -D(x;\bm{\lambda})e^{-\partial}\bigr)\varphi(x;\bm{\lambda}),
\end{align}
and their action on the polynomials is
\begin{align}
  &\mathcal{F}(\bm{\lambda})P_n(\eta(x;\bm{\lambda});\bm{\lambda})
  =f_n(\bm{\lambda})
  P_{n-1}(\eta(x;\bm{\lambda}+\bm{\delta});\bm{\lambda}+\bm{\delta}),\\
  &\mathcal{B}(\bm{\lambda})
  P_{n-1}(\eta(x;\bm{\lambda}+\bm{\delta});\bm{\lambda}+\bm{\delta})
  =b_{n-1}(\bm{\lambda})P_n(\eta(x;\bm{\lambda});\bm{\lambda}).
\end{align}
For the definition of the auxiliary function $\varphi(x;\bm{\lambda})$ see
eqs.\,(4.12) and (4.23) in \cite{os12}. Their explicit forms are also given
in \S\,5 of \cite{os12}.

The orthogonality relation is
\begin{equation}
  \sum_{x=0}^{x_{\text{max}}}\phi_0(x;\bm{\lambda})^2
  P_n(\eta(x;\bm{\lambda});\bm{\lambda})P_m(\eta(x;\bm{\lambda});\bm{\lambda})
  =\frac{1}{d_n(\bm{\lambda})^2}\,\delta_{nm}\quad
  (n,m=0,1,\ldots,n_{\text{max}}),
\end{equation}
where the explicit forms of the normalisation constants $d_n(\bm{\lambda})$
are given in \cite{os12}.

The coefficient of the leading term $c_n(\bm{\lambda})$, which appears in
$P_n(y;\bm{\lambda})=c_n(\bm{\lambda})P_n^{\text{monic}}(y;\bm{\lambda})$,
is given by
\begin{equation}
  c_n(\bm{\lambda})=\frac{(-1)^n\kappa^{-\frac12n(n-1)}}
  {\prod_{j=1}^n\eta(j;\bm{\lambda})}
  \prod_{j=0}^{n-1}\frac{\mathcal{E}(n;\bm{\lambda})
  -\mathcal{E}(j;\bm{\lambda})}
  {B(0;\bm{\lambda}+j\bm{\delta})},
\end{equation}
for all the examples in \cite{os12}.
Especially $n=1$ gives the relation
\begin{equation}
  \frac{B(0;\bm{\lambda})}{\mathcal{E}(1;\bm{\lambda})}
  =\frac{-c_1(\bm{\lambda})^{-1}}{\eta(1;\bm{\lambda})},
\end{equation}
which expresses the important relation among the basic quantities
$B(0;\bm{\lambda})$, $\mathcal{E}(1;\bm{\lambda})$, $\eta(1;\bm{\lambda})$
and $c_1(\bm{\lambda})$, see eq.\,(4.55) of \cite{os12}.
The Casorati determinant \eqref{ppolydef} becomes
\begin{align}
  &\quad\varphi_m(x;\bm{\lambda})^{-1}\,
  \text{W}[\check{P}_{n_1},\ldots,\check{P}_{n_m}](x;\bm{\lambda})\n
  &=\prod_{k=0}^{m-1}c_k(\bm{\lambda})\,\cdot
  \frac{\prod_{1\leq j<k\leq m}
  (\mathcal{E}(n_k;\bm{\lambda})-\mathcal{E}(n_j;\bm{\lambda}))}
  {\prod_{0\leq j<k\leq m-1}
  (\mathcal{E}(k;\bm{\lambda})-\mathcal{E}(j;\bm{\lambda}))}\,\cdot
  \prod_{k=1}^{m-1}\prod_{j=1}^k\eta(j;\bm{\lambda})\cdot
  \check{P}_{(n_1,\ldots,n_m)}(x;\bm{\lambda})\n
  &=(-1)^{\genfrac{(}{)}{0pt}{}{m}{2}}\kappa^{-\genfrac{(}{)}{0pt}{}{m}{3}}
  \prod_{1\leq j<k\leq m}
  \frac{\mathcal{E}(n_k;\bm{\lambda})-\mathcal{E}(n_j;\bm{\lambda})}
  {B(0;\bm{\lambda}+(j-1)\bm{\delta})}\cdot
  \check{P}_{(n_1,\ldots,n_m)}(x;\bm{\lambda}),
  \label{WP}
\end{align}
for all the examples in \cite{os12}.
Here $\check{P}_{(n_1,\ldots,n_m)}(x;\bm{\lambda})$ is a polynomial of
degree $\sum_{k=1}^mn_k-\frac12m(m-1)$ in
$\eta(x;\bm{\lambda}+(m-1)\bm{\delta})$ and satisfies the normalisation
\begin{equation}
  \check{P}_{(n_1,\ldots,n_m)}(0;\bm{\lambda})=1.
  \label{WP01}
\end{equation}

\subsection{The $(\phi_1,\ldots,\phi_{\ell})$-deleted systems}
\label{sec:12..l_delete}

For this very special case, the potential functions
\eqref{Bbarp}--\eqref{Dbarp}, the Hamiltonian
\eqref{dbfQMHb1bsdef}--\eqref{dbfQMAAd} and the Schr\"{o}dinger equation
\eqref{dbfQMHb1bnphi=..} of the modified system are:
\begin{align}
  &\!\!B_{\ell}(x;\bm{\lambda})=\bar{B}(x;\bm{\lambda})
  =B(x+\ell;\bm{\lambda})
  \frac{\text{W}[\check{P}_1,\ldots,\check{P}_{\ell}](x;\bm{\lambda})}
  {\text{W}[\check{P}_1,\ldots,\check{P}_{\ell}](x+1;\bm{\lambda})}
  \frac{\text{W}[\check{P}_1,\ldots,\check{P}_{\ell},\check{P}_0]
  (x+1;\bm{\lambda})}
  {\text{W}[\check{P}_1,\ldots,\check{P}_{\ell},\check{P}_0](x;\bm{\lambda})},
  \!\!\label{Bl}\\
  &D_{\ell}(x;\bm{\lambda})=\bar{D}(x;\bm{\lambda})
  =D(x;\bm{\lambda})
  \frac{\text{W}[\check{P}_1,\ldots,\check{P}_{\ell}](x+1;\bm{\lambda})}
  {\text{W}[\check{P}_1,\ldots,\check{P}_{\ell}](x;\bm{\lambda})}
  \frac{\text{W}[\check{P}_1,\ldots,\check{P}_{\ell},\check{P}_0]
  (x-1;\bm{\lambda})}
  {\text{W}[\check{P}_1,\ldots,\check{P}_{\ell},\check{P}_0](x;\bm{\lambda})},
  \label{Dl}\\
  &\mathcal{A}_{\ell}(\bm{\lambda})=\bar{\mathcal{A}}(\bm{\lambda})
  =\sqrt{B_{\ell}(x;\bm{\lambda})}
  -e^{\partial}\sqrt{D_{\ell}(x;\bm{\lambda})},\n
  &\mathcal{A}_{\ell}(\bm{\lambda})^{\dagger}
  =\bar{\mathcal{A}}(\bm{\lambda})^{\dagger}
  =\sqrt{B_{\ell}(x;\bm{\lambda})}
  -\sqrt{D_{\ell}(x;\bm{\lambda})}\,e^{-\partial},\\
  &\mathcal{H}_{\ell}(\bm{\lambda})
  =\bar{\mathcal{H}}(\bm{\lambda})
  =\mathcal{A}_{\ell}(\bm{\lambda})^{\dagger}\mathcal{A}_{\ell}(\bm{\lambda}),
  \quad
  x_{\text{max}}^{\ell}\eqdef N-\ell \text{ or } \infty,\\
  &\mathcal{H}_{\ell}(\bm{\lambda})\phi_{\ell,n}(x;\bm{\lambda})
  =\mathcal{E}(n;\bm{\lambda})\phi_{\ell,n}(x;\bm{\lambda})
  \quad(n=0,\ell+1,\ell+2,\ldots,n_{\text{max}}).
\end{align}
Note that the Hamiltonian $\mathcal{H}_{\ell}$ is an
$(x_{\text{max}}^{\ell}+1)\times(x_{\text{max}}^{\ell}+1)$ matrix
$\mathcal{H}_{\ell}=(\mathcal{H}_{\ell;x,y})$
$(x,y=0,1,\ldots,x_{\text{max}}^{\ell}$).
The explicit expression of $D_{\ell}(x;\bm{\lambda})$ can be analytically
continued and gives $D_{\ell}(0;\bm{\lambda})=0$.
We have $B_{\ell}(x_{\text{max}}^{\ell})=0$ for the finite case.
{}From \eqref{phibarnBp} we have
\begin{align}
  &\phi_{\ell,n}(x;\bm{\lambda})=\bar{\phi}_n(x;\bm{\lambda})
  =(-1)^{\ell}\phi_0(x;\bm{\lambda})
  \prod_{k=0}^{\ell-1}\sqrt{B(x+k;\bm{\lambda})}\n
  &\phantom{\phi_{\ell,0}(x;\bm{\lambda})=\bar{\phi}_0(x;\bm{\lambda})=}
  \times
  \frac{\text{W}[\check{P}_1,\ldots,\check{P}_{\ell},\check{P}_n]
  (x;\bm{\lambda})}
  {\sqrt{\text{W}[\check{P}_1,\ldots,\check{P}_{\ell}](x;\bm{\lambda})
  \text{W}[\check{P}_1,\ldots,\check{P}_{\ell}](x+1;\bm{\lambda})}}.
  \label{phil0}
\end{align}

Let us introduce the deforming polynomial
$\check{\xi}_{\ell}(x;\bm{\lambda})\eqdef
\xi_{\ell}(\eta(x;\bm{\lambda}+(\ell-1)\bm{\delta});\bm{\lambda})$,
which is a polynomial of degree $\ell$ in
$\eta(x;\bm{\lambda}+(\ell-1)\bm{\delta})$, and the deformed polynomial
$\check{P}_{\ell,n}(x;\bm{\lambda})\eqdef
P_{\ell,n}(\eta(x;\bm{\lambda}+\ell\bm{\delta});\bm{\lambda})$
($n=0,\ell+1,\ell+2,\ldots$), which is a polynomial of degree $n$ in
$\eta(x;\bm{\lambda}+\ell\bm{\delta})$, in terms of \eqref{WP} as follows:
\begin{equation}
  \check{\xi}_{\ell}(x;\bm{\lambda})\eqdef
  \check{P}_{(1,2,\ldots,\ell)}(x;\bm{\lambda}),\quad
  \check{P}_{\ell,n}(x;\bm{\lambda})\eqdef
  \check{P}_{(1,2,\ldots,\ell,n)}(x;\bm{\lambda}).
  \label{xilplndef}
\end{equation}
Then we have
\begin{equation}
  \check{\xi}_{\ell}(0;\bm{\lambda})=1,\quad
  \check{P}_{\ell,n}(0;\bm{\lambda})=1.
\end{equation}
We set $\check{P}_{\ell,n}(x;\bm{\lambda})=0$ for $n=1,\ldots,\ell$.
For the $q$-Racah case, which is the most generic case, the deforming
polynomial $\check{\xi}_{\ell}(x;\bm{\lambda})$ has the following form,
\begin{equation}
  \text{$q$-Racah : }
  \check{\xi}_{\ell}(x;\bm{\lambda})=
  \check{P}_{\ell}(-x;\mathfrak{t}(\bm{\lambda}+(\ell-1)\bm{\delta})),\quad
  \mathfrak{t}(\bm{\lambda})\eqdef-\bm{\lambda},
\end{equation}
and this is indeed a polynomial in $\eta(x;\bm{\lambda}+(\ell-1)\bm{\delta})$
because $\eta(-x;\bm{-\lambda}-(\ell-1)\bm{\delta})
=\eta(x;\bm{\lambda}+(\ell-1)\bm{\delta})(dq^{\ell-1})^{-1}$.
For the other cases the deforming polynomials are obtained from this in
certain limits. However these limits are not so trivial. So we present the
explicit forms of $\check{\xi}_{\ell}(x;\bm{\lambda})$ in \ref{sec:xil}.
Note that the deforming polynomial $\xi_{\ell}$ is related to
$P_{\ell,n}$ as
\begin{equation}
  \check{\xi}_{\ell}(x;\bm{\lambda})
  =\check{P}_{\ell-1,\ell}(x;\bm{\lambda}),
\end{equation}
and satisfies the recurrence relation:
\begin{align}
  B(0;\bm{\lambda}+\ell\bm{\delta})\check{\xi}_{\ell+1}(x;\bm{\lambda})
  &=B(x;\bm{\lambda}+\ell\bm{\delta})\varphi(x;\bm{\lambda}+\ell\bm{\delta})
  \check{\xi}_{\ell}(x;\bm{\lambda})\n
  &\quad
  -D(x;\bm{\lambda}+\ell\bm{\delta})\varphi(x-1;\bm{\lambda}+\ell\bm{\delta})
  \check{\xi}_{\ell}(x+1;\bm{\lambda}).
\end{align}

By using these quantities, eqs.\,\eqref{Bl}, \eqref{Dl} and \eqref{phil0} become
\begin{align}
  &B_{\ell}(x;\bm{\lambda})=
  \kappa^{\ell}B(x;\bm{\lambda}+\ell\bm{\delta})
  \frac{\check{\xi}_{\ell}(x;\bm{\lambda})}
  {\check{\xi}_{\ell}(x+1;\bm{\lambda})},
  \label{Blform}\\
  &D_{\ell}(x;\bm{\lambda})=
  \kappa^{\ell}D(x;\bm{\lambda}+\ell\bm{\delta})
  \frac{\check{\xi}_{\ell}(x+1;\bm{\lambda})}
  {\check{\xi}_{\ell}(x;\bm{\lambda})},
   \label{Dlform}\\
  &\phi_{\ell,0}(x;\bm{\lambda})=
  \frac{C(\ell,\bm{\lambda})}{\sqrt{B(0;\bm{\lambda}+\ell\bm{\delta})}}
  \frac{\phi_0(x;\bm{\lambda}+\ell\bm{\delta})}
  {\sqrt{\check{\xi}_{\ell}(x;\bm{\lambda})
  \check{\xi}_{\ell}(x+1;\bm{\lambda})}},
   \label{phil0form}\\
  &\phi_{\ell,n}(x;\bm{\lambda})
  =\phi_{\ell,0}(x;\bm{\lambda})(-1)^{\ell}
  \prod_{j=1}^{\ell}\frac{\mathcal{E}(n;\bm{\lambda})
  -\mathcal{E}(j;\bm{\lambda})}{\mathcal{E}(j;\bm{\lambda})}\cdot
  \check{P}_{\ell,n}(x;\bm{\lambda}),
  \label{philnform}
\end{align}
where $C(\ell,\bm{\lambda})$ is defined by
\begin{equation}
  C(\ell,\bm{\lambda})\eqdef
  \sqrt{B(0;\bm{\lambda}+\ell\bm{\delta})}\,(-1)^{\ell}
  \kappa^{-\frac14\ell(\ell-1)}
  \prod_{j=1}^{\ell}\frac{\mathcal{E}(j;\bm{\lambda})}
  {\sqrt{B(0;\bm{\lambda}+(j-1)\bm{\delta})}}.
\end{equation}
For {\em even} $\ell$, the deforming polynomial
$\check{\xi}_{\ell}(x;\bm{\lambda})
=\xi_{\ell}(\eta(x;\bm{\lambda}+(\ell-1)\bm{\delta});\bm{\lambda})$
is positive at integer points $x=0,1,\ldots,x_{\text{max}}^{\ell}+1$.
It has either no zero in the interval $0\leq x\leq x_{\text{max}}^{\ell}+1$
or when
it has zeros, an even number of zeros appear between two contiguous integers.
The deformed polynomial $P_{\ell,n}(y;\bm{\lambda})$ ($n\geq\ell+1$)
has $n-\ell$ zeros in the interval
$0\leq y\leq\eta(x_{\text{max}}^{\ell};\bm{\lambda}+\ell\bm{\delta})$ 
for even $\ell$.
These we have verified by numerical calculation for lower $\ell$ for
all the examples in \cite{os12}. To the best of our knowledge, no general
proof of this property has been reported.
Note that the normalisation of $\xi_{\ell}$ does not affect
$\mathcal{H}_{\ell}$. This system is not shape invariant.
We have not chosen the normalisation like as $\phi_0(0;\bm{\lambda})=1$,
namely
$\phi_{\ell,0}(0;\bm{\lambda})
=\frac{C(\ell,\bm{\lambda})}{\sqrt{B(0;\bm{\lambda}+\ell\bm{\delta})}
\sqrt{\check{\xi}_{\ell}(1;\bm{\lambda})}}\neq 1$.

The operators $\mathcal{A}_{\ell}(\bm{\lambda})$ and
$\mathcal{A}_{\ell}(\bm{\lambda})^{\dagger}$ connect the modified system
$\mathcal{H}_{\ell}(\bm{\lambda})=\mathcal{A}_{\ell}(\bm{\lambda})^{\dagger}
\mathcal{A}_{\ell}(\bm{\lambda})$
to the shape invariant system
$\mathcal{H}_{\ell}'(\bm{\lambda})=\mathcal{A}_{\ell}(\bm{\lambda})
\mathcal{A}_{\ell}(\bm{\lambda})^{\dagger}=\kappa^{\ell+1}\mathcal{H}
(\bm{\lambda}+(\ell+1)\bm{\delta})+\mathcal{E}(\ell+1;\bm{\lambda})
=\mathcal{H}^{[\ell+1]}(\bm{\lambda})$,
which is denoted by the rightmost vertical line in Fig.\,3.
The $n$-th level ($n\geq\ell+1$) of the modified system $\mathcal{H}_{\ell}$
is {\em iso-spectral\/} with the $n-\ell-1$-th level of the new
shape invariant system $\mathcal{H}_{\ell}'$:
\begin{align}
  &\mathcal{A}_{\ell}(\bm{\lambda})\phi_{\ell,n}(x;\bm{\lambda})
  =f_{\ell,n}(\bm{\lambda})
  \phi_{n-\ell-1}\bigl(x;\bm{\lambda}+(\ell+1)\bm{\delta}\bigr),\\
  &\mathcal{A}_{\ell}(\bm{\lambda})^{\dagger}
  \phi_{n-\ell-1}\bigl(x;\bm{\lambda}+(\ell+1)\bm{\delta}\bigr)
  =b_{\ell,n-1}(\bm{\lambda})\phi_{\ell,n}(x;\bm{\lambda}).
\end{align}
Here, $f_{\ell,n}(\bm{\lambda})$ and $b_{\ell,n-1}(\bm{\lambda})$ are the
factors of the energy eigenvalue, $\mathcal{E}(n;\bm{\lambda})
\!=\!f_{\ell,n}(\bm{\lambda})b_{\ell,n-1}(\bm{\lambda})$, and are defined by
\begin{equation}
  f_{\ell,n}(\bm{\lambda})\eqdef f_n(\bm{\lambda})
  \frac{\kappa^{-\frac{\ell}{2}}C(\ell,\bm{\lambda})}
  {B(0;\bm{\lambda}+\ell\bm{\delta})},\quad
  b_{\ell,n-1}(\bm{\lambda})\eqdef b_{n-1}(\bm{\lambda})
  \frac{B(0;\bm{\lambda}+\ell\bm{\delta})}
  {\kappa^{-\frac{\ell}{2}}C(\ell,\bm{\lambda})}.
\end{equation}
Note that $\mathcal{E}(n;\bm{\lambda})=\kappa^{\ell+1}
\mathcal{E}(n-\ell-1;\bm{\lambda}+(\ell+1)\bm{\delta})
+\mathcal{E}(\ell+1;\bm{\lambda})$ for $n\geq\ell+1$.
The forward and backward shift operators $\mathcal{F}_{\ell}(\bm{\lambda})$
and $\mathcal{B}_{\ell}(\bm{\lambda})$ and the similarity transformed
Hamiltonian $\widetilde{\mathcal{H}}_{\ell}(\bm{\lambda})$, which act on
the polynomial eigenfunctions, are defined by:
\begin{align}
  \mathcal{F}_{\ell}(\bm{\lambda})&\eqdef
  \phi_0\bigl(x;\bm{\lambda}+(\ell+1)\bm{\delta}\bigr)^{-1}\circ
  \mathcal{A}_{\ell}(\bm{\lambda})\circ\phi_{\ell,0}(x;\bm{\lambda})\n
  &=\frac{\kappa^{\frac{\ell}{2}}C(\ell,\bm{\lambda})}
  {\varphi(x;\bm{\lambda}+\ell\bm{\delta})
  \check{\xi}_{\ell}(x+1;\bm{\lambda})}
  \bigl(1-e^{\partial}\bigr),\\
  \mathcal{B}_{\ell}(\bm{\lambda})&\eqdef
  \phi_{\ell,0}(x;\bm{\lambda})^{-1}\circ
  \mathcal{A}_{\ell}(\bm{\lambda})^{\dagger}
  \circ\phi_0\bigl(x;\bm{\lambda}+(\ell+1)\bm{\delta}\bigr)
  \label{actBl}\\
  &=\frac{\kappa^{\frac{\ell}{2}}}{C(\ell,\bm{\lambda})}
  \Bigl(B(x;\bm{\lambda}+\ell\bm{\delta})\check{\xi}_{\ell}(x;\bm{\lambda})
  -D(x;\bm{\lambda}+\ell\bm{\delta})\check{\xi}_{\ell}(x+1;\bm{\lambda})
  e^{-\partial}\Bigl)
  \varphi(x;\bm{\lambda}+\ell\bm{\delta}),\n
  \widetilde{\mathcal{H}}_{\ell}(\bm{\lambda})&\eqdef
  \phi_{\ell,0}(x;\bm{\lambda})^{-1}\circ
  \mathcal{H}_{\ell}(\bm{\lambda})\circ\phi_{\ell,0}(x;\bm{\lambda})
  =\mathcal{B}_{\ell}(\bm{\lambda})\mathcal{F}_{\ell}(\bm{\lambda})\\
  &=\kappa^{\ell}\Bigl(
  B(x;\bm{\lambda}+\ell\bm{\delta})
  \frac{\check{\xi}_{\ell}(x;\bm{\lambda})}
  {\check{\xi}_{\ell}(x+1;\bm{\lambda})}(1-e^{\partial})
  +D(x;\bm{\lambda}+\ell\bm{\delta})
  \frac{\check{\xi}_{\ell}(x+1;\bm{\lambda})}
  {\check{\xi}_{\ell}(x;\bm{\lambda})}(1-e^{-\partial})\Bigr).\nonumber
\end{align}
Their action on the polynomials is
\begin{align}
  &\mathcal{F}_{\ell}(\bm{\lambda})\check{P}_{\ell,n}(x;\bm{\lambda})
  =f_{\ell,n}(\bm{\lambda})
  \check{P}_{n-\ell-1}(x;\bm{\lambda}+(\ell+1)\bm{\delta}),\\
  &\mathcal{B}_{\ell}(\bm{\lambda})
  \check{P}_{n-\ell-1}(x;\bm{\lambda}+(\ell+1)\bm{\delta})
  =b_{\ell,n-1}(\bm{\lambda})\check{P}_{\ell,n}(x;\bm{\lambda}),
  \label{dQMback}\\
  &\widetilde{\mathcal{H}}_{\ell}(\bm{\lambda})
  \check{P}_{\ell,n}(x;\bm{\lambda})
  =\mathcal{E}(n;\bm{\lambda})\check{P}_{\ell,n}(x;\bm{\lambda})
  \quad(n=0,\ell+1,\ell+2,\ldots,n_{\text{max}}).
\end{align}
For $n\geq\ell+1$, the above formula \eqref{actBl} provides a simple
expression of the modified eigenpolynomial $P_{\ell,n}$ in terms
of $\xi_{\ell}$ and the original eigenpolynomial $P_n$:
\begin{align}
  &\quad \kappa^{-\frac{\ell}{2}}C(\ell,\bm{\lambda})
  b_{\ell,n-1}(\bm{\lambda})\check{P}_{\ell,n}(x;\bm{\lambda})\n
  &=B(x;\bm{\lambda}+\ell\bm{\delta})\check{\xi}_{\ell}(x;\bm{\lambda})
  \varphi(x;\bm{\lambda}+\ell\bm{\delta})
  \check{P}_{n-\ell-1}(x;\bm{\lambda}+(\ell+1)\bm{\delta})\n
  &\quad
  -D(x;\bm{\lambda}+\ell\bm{\delta})\check{\xi}_{\ell}(x+1;\bm{\lambda})
  \varphi(x-1;\bm{\lambda}+\ell\bm{\delta})
  \check{P}_{n-\ell-1}(x-1;\bm{\lambda}+(\ell+1)\bm{\delta}).
  \label{plnform}
\end{align}

The orthogonality relation is ($x_{\text{max}}^{\ell}=N-\ell$ or $\infty$
and $n_{\text{max}}=N$ or $\infty$):
\begin{equation}
  \sum_{x=0}^{x_{\text{max}}^{\ell}}\phi_{\ell,0}(x;\bm{\lambda})^2
  \check{P}_{\ell,n}(x;\bm{\lambda})
  \check{P}_{\ell,m}(x;\bm{\lambda})
  =\frac{1}{d_{\ell,n}(\bm{\lambda})^2}\,\delta_{nm}\quad
  (n,m=0,\ell+1,\ell+2,\ldots,n_{\text{max}}).
\end{equation}
Here the normalisation constant $d_{\ell,n}(\bm{\lambda})$ is
\begin{equation}
  d_{\ell,n}(\bm{\lambda})^2=d_n(\bm{\lambda})^2
  \prod_{j=1}^{\ell}
  \frac{\mathcal{E}(n;\bm{\lambda})-\mathcal{E}(j;\bm{\lambda})}
  {\mathcal{E}(j;\bm{\lambda})^2},
\end{equation}
which is a consequence of \eqref{orthocPcP} and positive for even $\ell$.
The weight function $\phi_{\ell,0}(x;\bm{\lambda})^2$ is positive definite
for even $\ell$.

\subsection{Explicit forms of the deforming polynomial $\xi_{\ell}$}
\label{sec:xil}

As shown in the preceding subsection, the various quantities, the functions
$B_{\ell}(x)$ \eqref{Blform}, $D_{\ell}(x)$ \eqref{Dlform}, which specify
the Hamiltonian $\mathcal{H}_{\ell}$ and the eigenfunctions $\phi_{\ell,0}$
\eqref{phil0form}, $\phi_{\ell,n}(x)$ \eqref{philnform} are determined
by the {\em deforming polynomial} $\check{\xi}_{\ell}(x;\bm{\lambda})
=\xi_{\ell}(\eta(x;\bm{\lambda}+(\ell-1)\bm{\delta});\bm{\lambda})$.
This deforming polynomial $\check{\xi}_{\ell}(x;\bm{\lambda})$ is positive at
integer points $x=0,1,\ldots,x_{\text{max}}^{\ell}$ for {\em even} $\ell$.
Here we give the list of the explicit forms of the deforming polynomial
$\check{\xi}_{\ell}(x;\bm{\lambda})$.
The list contains the name of the polynomial ({\em e.g.}\ Racah) and the
corresponding section number of \cite{os12} together with that of
Koekoek-Swarttouw's review \cite{koeswart}, the definition of the
{\em twist operator} $\mathfrak{t}$ and the form of the deforming polynomial
$\check{\xi}_{\ell}(x;\bm{\lambda})$ in terms of the original polynomial
$\check{P}_{\ell}(x;\bm{\lambda})$ (the Racah polynomial):
\begin{align}
  & \text{name (\S\ of \cite{os12}, \S\ of \cite{koeswart}):}\n
  &\text{Racah (5.1.1, 1.2):}\qquad
  \mathfrak{t}(\bm{\lambda})\eqdef-\bm{\lambda},\n
  &\qquad\check{\xi}_{\ell}(x;\bm{\lambda})=
  \check{P}_{\ell}(-x;\mathfrak{t}(\bm{\lambda}+(\ell-1)\bm{\delta})),\\
  &\text{Hahn (5.1.2, 1.5):}\qquad
  \mathfrak{t}(\bm{\lambda})\eqdef-\bm{\lambda},\n
  &\qquad\check{\xi}_{\ell}(x;\bm{\lambda})=
  \check{P}_{\ell}(-x;\mathfrak{t}(\bm{\lambda}+(\ell-1)\bm{\delta})),\\
  &\text{dual Hahn (5.1.3, 1.6):}\qquad
  \mathfrak{t}(\bm{\lambda})\eqdef-\bm{\lambda},\n
  &\qquad\check{\xi}_{\ell}(x;\bm{\lambda})=
  \check{P}_{\ell}(-x;\mathfrak{t}(\bm{\lambda}+(\ell-1)\bm{\delta})+(0,2,0)),
  \\
  &\text{Krawtchouk (5.1.4, 1.10):}\qquad
  \mathfrak{t}(\bm{\lambda})\eqdef(p,-N),\n
  &\qquad\check{\xi}_{\ell}(x;\bm{\lambda})=
  \check{P}_{\ell}(-x;\mathfrak{t}(\bm{\lambda}+(\ell-1)\bm{\delta})),\\
  &\text{$q$-Racah (5.1.5, 3.2):}\qquad
  \mathfrak{t}(\bm{\lambda})\eqdef-\bm{\lambda},\n
  &\qquad\check{\xi}_{\ell}(x;\bm{\lambda})=
  \check{P}_{\ell}(-x;\mathfrak{t}(\bm{\lambda}+(\ell-1)\bm{\delta})),\\
  &\text{$q$-Hahn (5.1.6, 3.6):}\qquad
  \mathfrak{t}(\bm{\lambda})\eqdef-(b,a,N),\n
  &\qquad\check{\xi}_{\ell}(x;\bm{\lambda})=
  \check{P}_{\ell}(x-N+\ell-1;\mathfrak{t}(\bm{\lambda}+(\ell-1)\bm{\delta}))
  \frac{(-1)^{\ell}a^{\ell}q^{\frac12\ell(\ell-1)}(b;q)_{\ell}}
  {(a;q)_{\ell}},\\
  &\text{dual $q$-Hahn (5.1.7, 3.7):}\n
  &\qquad\check{\xi}_{\ell}(x;\bm{\lambda})=
   {}_3\phi_2\Bigl(
  \genfrac{}{}{0pt}{}{q^{-\ell},q^x,a^{-1}b^{-1}q^{-x+2-\ell}}
  {a^{-1}q^{-\ell+1},q^{N-\ell+1}}\Bigm|q\,;bq^N\Bigr),\\
  &\text{quantum $q$-Krawtchouk (5.1.8, 3.14):}\n
  &\qquad\check{\xi}_{\ell}(x;\bm{\lambda})=
   {}_3\phi_2\Bigl(
  \genfrac{}{}{0pt}{}{q^{-\ell},0,q^{-x+N-\ell+1}}
  {p^{-1}q^{-\ell},q^{N-\ell+1}}\Bigm|q\,;q\Bigr)(pq\,;q)_{\ell},\\
  &\text{dual quantum $q$-Krawtchouk (5.1.8, ---):}\n
  &\qquad\check{\xi}_{\ell}(x;\bm{\lambda})=
   {}_2\phi_1\Bigl(
  \genfrac{}{}{0pt}{}{q^{-\ell},q^x}
  {q^{N-\ell+1}}\Bigm|q\,;pq^{N+1}\Bigr),\\
  &\text{$q$-Krawtchouk (5.1.9, 3.15):}\qquad
  \mathfrak{t}(\bm{\lambda})\eqdef-\bm{\lambda},\n
  &\qquad\check{\xi}_{\ell}(x;\bm{\lambda})=
  \check{P}_{\ell}(x-N+\ell-1;
  \mathfrak{t}(\bm{\lambda}+(\ell-1)\bm{\delta})+(-2,0))
  (-1)^{\ell}q^{\ell^2}p^{\ell},\\
  &\text{dual $q$-Krawtchouk
 (in the standard parametrization) (5.1.9, 3.17):}\n
  &\qquad\check{\xi}_{\ell}(x;\bm{\lambda})=
  {}_3\phi_1\Bigl(
  \genfrac{}{}{0pt}{}{q^{-\ell},q^x,c^{-1}q^{-x+N-\ell+1}}
  {q^{N-\ell+1}}\Bigm|q\,;cq^{\ell}\Bigr),\\
  &\text{affine $q$-Krawtchouk (5.1.10, 3.16):}\n
  &\qquad\check{\xi}_{\ell}(x;\bm{\lambda})=
  {}_2\phi_1\Bigl(
  \genfrac{}{}{0pt}{}{q^{-\ell},q^{-x+N-\ell+1}}
  {q^{N-\ell+1}}\Bigm|q\,;p^{-1}\Bigr)
  \frac{(-1)^{\ell}p^{\ell}q^{\frac12\ell(\ell+1)}}{(pq\,;q)_{\ell}},\\
  &\text{alternative $q$-Hahn (5.3.1, ---):}\qquad
  \mathfrak{t}(\bm{\lambda})\eqdef-(b,a,N),\n
  &\qquad\check{\xi}_{\ell}(x;\bm{\lambda})=
  \check{P}_{\ell}(x-N+\ell-1;\mathfrak{t}(\bm{\lambda}+(\ell-1)\bm{\delta}))
  \frac{(-1)^{\ell}q^{-\frac12\ell(\ell-1)}(a;q)_{\ell}}
  {a^{\ell}(b;q)_{\ell}},\\
  &\text{alternative $q$-Krawtchouk (5.3.2, ---):}\qquad
  \mathfrak{t}(\bm{\lambda})\eqdef-\bm{\lambda},\n
  &\qquad\check{\xi}_{\ell}(x;\bm{\lambda})=
  \check{P}_{\ell}(x-N+\ell-1;
  \mathfrak{t}(\bm{\lambda}+(\ell-1)\bm{\delta})+(-2,0))
  (-1)^{\ell}p^{-\ell}q^{-\ell^2},\\
  &\text{alternative affine $q$-Krawtchouk (5.3.3, ---):}\n
  &\qquad\check{\xi}_{\ell}(x;\bm{\lambda})=
  {}_2\phi_1\Bigl(
  \genfrac{}{}{0pt}{}{q^{-\ell},q^{-x+N-\ell+1}}{q^{N-\ell+1}}
  \Bigm|q\,;pq^{x+\ell+1}\Bigr)
  \frac{1}{(pq\,;q)_{\ell}},\\
  &\text{Meixner (5.2.1, 1.9):}\qquad
  \mathfrak{t}(\bm{\lambda})\eqdef(-\beta,c),\n
  &\qquad\check{\xi}_{\ell}(x;\bm{\lambda})=
  \check{P}_{\ell}(-x;
  \mathfrak{t}(\bm{\lambda}+(\ell-1)\bm{\delta})),\\
  &\text{Charlier (5.2.2, 1.12):}\qquad
  \mathfrak{t}(\bm{\lambda})\eqdef-\bm{\lambda},\n
  &\qquad\check{\xi}_{\ell}(x;\bm{\lambda})=
  \check{P}_{\ell}(-x;\mathfrak{t}(\bm{\lambda}+(\ell-1)\bm{\delta})),\\
  &\text{little $q$-Jacobi (5.2.3, 3.12):}\qquad
  \mathfrak{t}(\bm{\lambda})\eqdef-\bm{\lambda},\n
  &\qquad\check{\xi}_{\ell}(x;\bm{\lambda})=
  \check{P}_{\ell}(x+b'+\ell;
  \mathfrak{t}(\bm{\lambda}+(\ell-1)\bm{\delta})-(2,2))
  a^{-\ell}b^{-\ell}q^{-\ell(\ell+1)},\quad b=q^{b'},\\
  &\text{dual little $q$-Jacobi (5.2.3, ---):}\n
  &\qquad\check{\xi}_{\ell}(x;\bm{\lambda})=
  {}_3\phi_2\Bigl(
  \genfrac{}{}{0pt}{}{q^{-\ell},q^x,a^{-1}b^{-1}q^{-x-\ell}}{b^{-1}q^{-\ell},0}
  \Bigm|q\,;q\Bigr),\\
  &\text{$q$-Meixner (5.2.4, 3.13):}\n
  &\qquad\check{\xi}_{\ell}(x;\bm{\lambda})=
  {}_2\phi_1\Bigl(
  \genfrac{}{}{0pt}{}{q^{-\ell},q^x}{b^{-1}q^{-\ell}}
  \Bigm|q\,;-b^{-1}c^{-1}q^{1-x}\Bigr),\\
\ignore{
  &\text{dual $q$-Meixner (5.2.4, ---):}\n
  &\qquad\check{\xi}_{\ell}(x;\bm{\lambda})=
  {}_2\phi_1\Bigl(
  \genfrac{}{}{0pt}{}{q^{-\ell},q^x}{b^{-1}q^{-\ell}}
  \Bigm|q\,;-b^{-1}c^{-1}\Bigr),\\
}
  &\text{little $q$-Laguerre/Wall (5.2.5, 3.20):}\n
  &\qquad\check{\xi}_{\ell}(x;\bm{\lambda})=
  {}_1\phi_1\Bigl(
  \genfrac{}{}{0pt}{}{q^{-\ell}}{a^{-1}q^{-\ell}}
  \Bigm|q\,;a^{-1}q^x\Bigr)
  (-1)^{\ell}a^{-\ell}q^{-\frac12\ell(\ell+1)}(aq\,;q)_{\ell},\\
  &\text{Al-Salam-Carlitz \Romannumeral{2} (5.2.6, 3.25):}\n
  &\qquad\check{\xi}_{\ell}(x;\bm{\lambda})=
  {}_2\phi_1\Bigl(
  \genfrac{}{}{0pt}{}{q^{-\ell},q^x}{0}
  \Bigm|q\,;a^{-1}q^{1-x}\Bigr),\\
  &\text{alternative $q$-Charlier (5.2.7, 3.22):}\n
  &\qquad\check{\xi}_{\ell}(x;\bm{\lambda})=
  {}_2\phi_0\Bigl(
  \genfrac{}{}{0pt}{}{q^{-\ell},-a^{-1}q^{-\ell}}{-}
  \Bigm|q\,;-aq^{x+2\ell}\Bigr)
  (-a)^{-\ell}q^{-\ell^2},\\
  &\text{dual alternative $q$-Charlier (5.2.7, ---):}\n
  &\qquad\check{\xi}_{\ell}(x;\bm{\lambda})=
  {}_3\phi_2\Bigl(
  \genfrac{}{}{0pt}{}{q^{-\ell},q^x,-a^{-1}q^{1-x-\ell}}{0,0}
  \Bigm|q\,;q\Bigr),\\
  &\text{$q$-Charlier (5.2.8, 3.23):}\n
  &\qquad\check{\xi}_{\ell}(x;\bm{\lambda})=
  {}_2\phi_0\Bigl(
  \genfrac{}{}{0pt}{}{q^{-\ell},q^x}{-}
  \Bigm|q\,;-a^{-1}q^{\ell+1-x}\Bigr).
\ignore{
  &\text{dual $q$-Charlier (5.2.8, ---):}\n
  &\qquad\check{\xi}_{\ell}(x;\bm{\lambda})=
  {}_2\phi_0\Bigl(
  \genfrac{}{}{0pt}{}{q^{-\ell},q^x}{-}
  \Bigm|q\,;-a^{-1}q^{\ell}\Bigr).
}
\end{align}
The twist operator $\mathfrak{t}$ is not listed when it is not used
for the definition of $\xi_{\ell}$.

\subsection{Supplementary data on dual orthogonal polynomials}
\label{sec:suppl}

Here we present the extra data for various dual orthogonal polynomials
which were not listed in \cite{os12}. They are the dual quantum
$q$-Krawtchouk in \S\,5.1.8, the dual little $q$-Jacobi in \S\,5.2.3,
the dual alternative $q$-Charlier in \S\,5.2.7 in \cite{os12}.\\
$\bullet$ dual quantum $q$-Krawtchouk:
\begin{align}
  &q^{\bm{\lambda}}=(p,q^N),\quad\bm{\delta}=(0,-1),\quad
  \kappa=q^{-1},\quad p>q^{-N},\\
  &\mathcal{E}(n;\bm{\lambda})=q^{-n}-1,\quad\eta(x;\bm{\lambda})=1-q^x,\\
  &P_n(\eta(x;\bm{\lambda});\bm{\lambda})=
  {}_2\phi_1\Bigl(
  \genfrac{}{}{0pt}{}{q^{-n},q^{-x}}{q^{-N}}
  \Bigm|q\,;pq^{x+1}\Bigr),\\
  &\phi_0(x;\bm{\lambda})^2
  =\frac{(q\,;q)_N}{(q\,;q)_x(q\,;q)_{N-x}}\,
  \frac{p^{-x}q^{-Nx}}{(p^{-1}q^{-x}\,;q)_x},\\
  &d_n(\bm{\lambda})^2
  =\frac{(q\,;q)_N}{(q\,;q)_n(q\,;q)_{N-n}}\,
  \frac{p^{-n}q^{n(n-1-N)}}{(p^{-1}q^{-N}\,;q)_n}
  \times(p^{-1}q^{-N}\,;q)_N,\\
  &R_1(z;\bm{\lambda})=(q^{-\frac12}-q^{\frac12})^2 z',\quad
  z'\eqdef z+1,\\
  &R_0(z;\bm{\lambda})=(q^{-\frac12}-q^{\frac12})^2
  z^{\prime\,2},\\
  &R_{-1}(z;\bm{\lambda})=(q^{-\frac12}-q^{\frac12})^2
  \bigl(-z^{\prime\,2}+p^{-1}(1+p+q^{-N-1})z'
  -p^{-1}q^{-N}(1+q^{-1})\bigr),\\
  &\varphi(x;\bm{\lambda})=q^x,\quad
  f_n(\bm{\lambda})=q^{-n}-1,\quad b_n(\bm{\lambda})=1.
\end{align}
$\bullet$ dual little $q$-Jacobi:
\begin{align}
  &q^{\bm{\lambda}}=(a,b),\quad\bm{\delta}=(0,1),\quad
  \kappa=q,\quad 0<a<q^{-1},\quad b<q^{-1},\\
  &\mathcal{E}(n;\bm{\lambda})=1-q^n,\quad
  \eta(x;\bm{\lambda})=(q^{-x}-1)(1-abq^{x+1}),\\
  &P_n(\eta(x;\bm{\lambda});\bm{\lambda})=
  {}_3\phi_1\Bigl(
  \genfrac{}{}{0pt}{}{q^{-n},q^{-x},abq^{x+1}}{bq}
  \Bigm|q\,;a^{-1}q^n\Bigr),\\
  &\phi_0(x;\bm{\lambda})^2
  =\frac{(bq,abq\,;q)_x\,a^xq^{x^2}}{(q,aq\,;q)_x}\,
  \frac{1-abq^{2x+1}}{1-abq},\\
  &d_n(\bm{\lambda})^2
  =\frac{(bq\,;q)_n}{(q\,;q)_n}(aq)^n
  \times\frac{(aq\,;q)_{\infty}}{(abq^2;q)_{\infty}},\\
  &R_1(z;\bm{\lambda})=(q^{-\frac12}-q^{\frac12})^2 z',\quad
  z'\eqdef z-1,\\
  &R_0(z;\bm{\lambda})=(q^{-\frac12}-q^{\frac12})^2z^{\prime\,2},\\
  &R_{-1}(z;\bm{\lambda})=(q^{-\frac12}-q^{\frac12})^2
  \bigl((1+abq)z^{\prime\,2}+(1+a)z'\bigr),\\
  &\varphi(x;\bm{\lambda})=\frac{q^{-x}-abq^{x+2}}{1-abq^2},\quad
  f_n(\bm{\lambda})=1-q^n,\quad b_n(\bm{\lambda})=1.
\end{align}
$\bullet$ dual alternative $q$-Charlier:
\begin{align}
  &q^{\bm{\lambda}}=a,\quad\bm{\delta}=1,\quad
  \kappa=q,\quad a>0,\\
  &\mathcal{E}(n;\bm{\lambda})=1-q^n,\quad
  \eta(x;\bm{\lambda})=(q^{-x}-1)(1+aq^x),\\
  &P_n(\eta(x;\bm{\lambda});\bm{\lambda})=q^{nx}
  {}_2\phi_1\Bigl(
  \genfrac{}{}{0pt}{}{q^{-n},q^{-x}}{0}
  \Bigm|q\,;-a^{-1}q^{1-x}\Bigr)
  ={}_3\phi_0\Bigl(
  \genfrac{}{}{0pt}{}{q^{-n},q^{-x},-aq^x}{-}
  \Bigm|q\,;-a^{-1}q^n\Bigr),\\
  &\phi_0(x;\bm{\lambda})^2
  =\frac{a^xq^{\frac12x(3x-1)}(-a\,;q)_x}{(q\,;q)_x}\,
  \frac{1+aq^{2x}}{1+a},\\
  &d_n(\bm{\lambda})^2
  =\frac{a^nq^{\frac12n(n+1)}}{(q\,;q)_n}
  \times\frac{1}{(-aq\,;q)_{\infty}},\\
  &R_1(z;\bm{\lambda})=(q^{-\frac12}-q^{\frac12})^2 z',\quad
  z'\eqdef z-1,\\
  &R_0(z;\bm{\lambda})=(q^{-\frac12}-q^{\frac12})^2z^{\prime\,2},\\
  &R_{-1}(z;\bm{\lambda})=(q^{-\frac12}-q^{\frac12})^2
  \bigl((1-a)z^{\prime\,2}+z'\bigr),\\
  &\varphi(x;\bm{\lambda})=\frac{q^{-x}+aq^{x+1}}{1+aq},\quad
  f_n(\bm{\lambda})=1-q^n,\quad b_n(\bm{\lambda})=1.
\end{align}
We note that the little $q$-Jacobi polynomial eq.\,(5.193) and
the alternative $q$-Charlier polynomial eq.\,(5.248) in \cite{os12}
can be rewritten as
\begin{alignat}{2}
  &\text{little $q$-Jacobi}:
  &P_n(\eta(x;\bm{\lambda});\bm{\lambda})=
  {}_3\phi_1\Bigl(
  \genfrac{}{}{0pt}{}{q^{-n},q^{-x},abq^{n+1}}{bq}
  \Bigm|q\,;a^{-1}q^x\Bigr),\\
  &\text{alternative $q$-Charlier}:\quad
  &P_n(\eta(x;\bm{\lambda});\bm{\lambda})=
  {}_3\phi_0\Bigl(
  \genfrac{}{}{0pt}{}{q^{-n},q^{-x},-aq^n}{-}
  \Bigm|q\,;-a^{-1}q^x\Bigr).
\end{alignat}


\end{document}